\documentclass[12pt]{article}
\pdfoutput=1
\usepackage{bm}
\usepackage{mathrsfs}
\usepackage{amsbsy}
\usepackage{graphicx}
\usepackage{subfigure}
\usepackage{amsmath}
\usepackage{amssymb}
\usepackage{color}
\usepackage[numbers,sort&compress]{natbib}
\newcommand{\be}{\begin{equation}}
\newcommand{\ee}{\end{equation}}

\numberwithin{equation}{section}
\begin{document}
	
	\begin{center}
		{\bf \Large{Page Curve and Phase Transition in deformed Jackiw-Teitelboim Gravity}}\\
		
		\vspace{1.6cm}
		{\small{\textbf{Cheng-Yuan Lu}$^{1}$,~\textbf{Ming-Hui Yu}$^{1}$, ~\textbf{Xian-Hui Ge}$^{1*}$,~\textbf{Li-Jun Tian}$^{1}$}}\let\thefootnote\relax
\footnotetext{* Corresponding author. gexh@shu.edu.cn}\\

		\vspace{0.8cm}
		
		$^1${\it Department of Physics, Shanghai University, Shanghai 200444,  China} \\

		\vspace{1.6cm}

		\begin{abstract}
		We consider the entanglement island in a deformed Jackiw-Teitelboim black hole in the presence of the phase transition. This black hole has the van der Waals-Maxwell-like phase structure as it is coupled with a Maxwell field. We study the behavior of the Page curve of this black hole by using the island paradigm. In the fixed charge ensemble, we discuss different situations with different charges that influence the system's phase structure. There is only a Hawking-Page phase transition in the absence of charges, which leads to an unstable small black hole. Hence, the related Page curve does not exist. However, a van der Waals-Maxwell-like phase transition occurs in the presence of charges. This yields three black hole solutions. The Page curve of the middle size black hole does not exist. For the extremal black hole, the Page time approaches zero in the phase transition situation but becomes divergent without the phase transition. In a word, we study the Page curve and the island paradigm for  different black hole phases and in  different phase transition situations.
		\end{abstract}
	\end{center}
\newpage
\tableofcontents
\newpage

\section{Introduction} \label{Introduction}
\qquad It is widely believed that studying the information loss paradox is significant for revealing quantum gravity. In 1975, Stephen Hawking discovered that the black hole would emit black-body radiation which does not carry any information about the interior of the black hole \cite{ref1,ref2}. If a black hole is formed by the pure state, then the whole process of black hole evaporation corresponds to the evolution from the pure state to the mixed state, which breaks the unitary principle of quantum mechanics. If quantum mechanics is respected, the fine-grained entropy, also called entanglement entropy or von-Neuman entropy,  of Hawking radiation should drop to zero at the end of the whole process. However, Don Page pointed out that if black hole evaporation is a unitary process, then the entanglement entropy of the radiation should linearly increase at the early time, and decrease after a special time -- Page time, this curve is called the Page curve \cite{PC1,PC2}. Recently, there is a great breakthrough in this issue. By using the island formula \cite{QES}, the Page curve emerges naturally. It leads to a decreasing fine-grained entropy for the Hawking radiation due to the emergence of the island in the black hole interior. The fine-grained entropy of Hawking radiation can be expressed as \cite{island rule,entanglement wedge}
\be
S_{\text{rad}}=\text{Min}_{I}\left\{\text{Ext}_{I}\left[\frac{\text{Area}(\partial I)}{4 G_{\mathrm{N}}}+S_{\mathrm{vN}}\left(R\cup I\right)\right]\right\},
\ee
where $I$ represents the island, $G_N$ is the Newton constant, $R$ is the radiation region and $\partial I$ is the boundary of the island. Crucially, the calculations include the contribution of islands inside the black hole to the radiation entropy. In order to obtain the right answer, we should first extremize the generalized entropy and then pick up the minimum value to match this formula. Before the Page time, there are no islands, so the generalized entropy without the island is the smaller one. However, the generalized entropy with the island has more weight in the process of minimization after the Page time. Therefore, the entanglement wedge not only includes the outgoing Hawking radiation, but also the black hole interior purifying partner.

The island formula was first proposed in a solvable two dimensional evaporating black hole model which is constructed by coupling a bath to a nearly-$\rm{AdS_2}$ black hole \cite{bulk entropy}. Different from the evaporating black hole, the island of the eternal black hole is well established and outside the event horizon  \cite{eternal}. Although the island formula is derived from Jackiw-Teitelboim(JT) black holes, its application is beyond the context of Anti-de Sitter($\rm{AdS}$) spacetime. It can be applied to other spacetime backgrounds, such as the other $\text{AdS}_2$ black hole \cite{JT1,JT2,JT3,JT4,JT5,JT6,JT7,JT8}, the two-dimensional(2D) dilaton gravity \cite{2D0,2D1,2D2,2D3,2D4,2D5,2D6}, the higher-dimensional spacetime \cite{high0,high1,high2,high3,high4,high5,high6,high7,HosseiniMansoori:2022hok,high8}, the cosmology \cite{yz1,yz2,yz3,yz4,yz5,yz6,yz7,yz8}, the boundary conformal field theory(BCFT) \cite{BCFT1,BCFT2,BCFT3,BCFT4,BCFT5,BCFT6,BCFT7}. Interestingly, there are also some related works for the Sachdev-Ye-Kitaev(SYK) model \cite{SYK1,SYK2,SYK3,SYK4,SYK5,SYK6}.

Up to now, most studies have not yet considered the influence of the black hole phase transition on the Page curve. Considering the possibility that the phase transition may change the structure of the Page curve, we would like to study the relationship among the island formula, the Page curve, and the phase transition. Based on this motivation, this paper studies a special 2D black hole solution that yields van der Waals-Maxwell-like phase structure in deformed JT gravity \cite{phase1,phase2,phase3}. Following \cite{eternal}, we consider a two-side eternal black hole coupled to flat thermal baths. Then we use these strategies to obtain the Page curve and discuss the effect of the phase transition on the Page curve.

This paper is organized as follows. In section 2, we discuss the black hole solution which is derived from JT-Maxwell gravity and introduce our toy model -- the two-sided deformed JT black hole coupled thermal baths. In section 3, we calculate the evolution of entanglement entropy without island and display the paradox. Then, we consider an island that would avoid the divergence of the entanglement entropy at late times. In section 4, based on the previous result, we plot the Page curve and derive the scrambling time. In section 5, we considered these results in combination with the phase transition. We discuss the change of the Page time with different charges in the canonical ensemble. We consider both the Hawking-Page phase transition without charges and the van der Waals-Maxwell-like phase transition with charges. Finally, we give a summary and some remarks in the last section. Hereafter,we use the unit $\hbar=k_B=c=1$.

\section{Set up}

\qquad In this section, we briefly review the deformed JT gravity with a Maxwell field \cite{d1,d2,d3,d4,phase1}. We can write down the action for the deformed JT gravity with metric $g$ and dilaton coupled to conformal matter, the action can be written as follows \cite{jt1,jt2},
\be
\begin{aligned}
S[g, \phi, \Phi] &=S_{\mathrm{top}}[g]+S_{\mathrm{JT}}[g, \phi]+S_{\rm{m}}[g, \Phi] \\
&=\frac{\phi_{0}}{16 \pi G_{\mathrm{N}}}\left[\int_{\mathcal{M}} \sqrt{-g} R+2 \int_{\partial \mathcal{M}} \sqrt{-\gamma} K\right]\\&+\frac{1}{16 \pi G_{\mathrm{N}}}\left[\int_{\mathcal{M}} \sqrt{-g} \phi(R+2)+2 \int_{\partial \mathcal{M}} \sqrt{-\gamma} \phi_{b}(K-1)\right]\\ &+S_{\rm{m}}[g, \Phi].
\end{aligned}
\ee

The first term $S_{\rm{top}}$ is a purely topological term and only contributes a constant $\phi_{0}\chi$, where $\chi$ is the Euler characteristic of the corresponding manifold $\mathcal{M}$. The second term $S_{\rm{JT}}$ is the consequence of dimensional reduction. The dilaton $\phi$, the Gibbons-Hawking-York boundary term, a counter-term with boundary metric $\gamma$ and curvature $K$, and boundary dilaton value $\phi_{\rm{b}}$ are included in $S_{\rm{JT}}$. The last term $S_{\rm{m}}$ is simply a 2D CFT action for a matter field, which does not couple to the dilaton directly.

Hereafter, we consider a deformed JT gravity plus a Maxwell field. The action for this theory reads as \cite{phase1}
\be
\begin{aligned}
S_{\mathrm{JT}}^{\text {charg }} &=\frac{\phi_{0}}{2}\left(\int_{\mathcal{M}} d^{2} x \sqrt{-g} R+2 \int_{\partial M} \sqrt{-\gamma} K d \tau\right) \\
&+\frac{1}{2} \int_{\mathcal{M}} d^{2} x \sqrt{-g}\left(\phi R+\frac{V(\phi)}{l^{2}}-\frac{1}{2} Z(\phi) \mathcal{F}^{2}\right) \\
&+\frac{1}{2} \int_{\partial \mathcal{M}} d \tau \sqrt{-\gamma} K_{\mu} Z(\phi) \mathcal{F}^{\mu \nu} \mathcal{A}_{\nu}+\int_{\partial \mathcal{M}} \sqrt{-\gamma} \phi_b K d \tau \\
&+\int_{\partial \mathcal{M}} \sqrt{-\gamma} \mathcal{L}_{\mathrm{ct}} d \tau,
\end{aligned}
\ee
where $l$ is the Ads radius, $\mathcal{F}$ is the Maxwell field and $\mathcal{L}_{\mathrm{ct}}$ is the counterterms. The first bracket is the topological term for JT gravity. Through the variation of this action, we can obtain the black hole solution of the above theory in the Schwarzschild coordinates,
\begin{subequations}
\be
\begin{aligned}
d s^{2} =-f(r) d t^{2}+\frac{d r^{2}}{f(r)}, \\
\end{aligned}
\label{s}
\ee
\be
\begin{aligned}
V(\phi) =2 \phi+\frac{a_{0}}{\phi^{\eta}},\quad \phi(r) =r,  \\
\end{aligned}
\ee
\be
\begin{aligned}
f(r) =\frac{r^{2}-r_{\rm{H}}^{2}}{l^{2}}+\frac{a_{0}}{(1-\eta) l^{2}}\left(r^{1-\eta}-r_{\rm{H}}^{1-\eta}\right)-\frac{Q^{2}}{(1-\zeta) l^{2}}\left(r^{1-\zeta}-r_{\rm{H}}^{1-\zeta}\right),   \\
\end{aligned}
\label{fr}
\ee
\be
\begin{aligned}
Z(\phi) =r^{\zeta}, \quad \mathcal{A}_{t}(r)=\mu+\frac{Q}{1-\zeta} r^{1-\zeta},
\end{aligned}
\ee
\end{subequations}
where $l$ is the AdS radius, $r_{\rm{H}}$ is the event horizon of the black hole, $Q$ represents the charge density, and the parameters satisfy $a_{0}>0,\eta>0,\zeta>0$. The dilaton potential $V\left(\phi\right)$ satisfies $V\left(\phi_{\infty}\right)=2r$ for $r\rightarrow\infty$. Note that $\mu=\frac{\mathcal{Q}}{\zeta-1} r_{\rm{H}}^{1-\zeta}$ is the chemical potential. The blacken factor $f(r)$ recovers the standard JT black hole for $f(r)=\frac{1}{l^{2}}\left(r^{2}-r_{\rm{H}}^{2}\right)$ as $a_{0}=Q^{2}=0$. The Hawking temperature and the entropy of the black hole are given by
\be
T_{\rm{H}}=\frac{f'\left(r_{\rm{H}}\right)}{4\pi}=\frac{2 r_{\rm{H}}^{\zeta+1}+a_{0} r_{\rm{H}}^{\zeta-\eta}-Q^{2}}{4 \pi r_{\rm{H}}^{\zeta}},
\label{T}
\ee
\be
S=2 \pi r_{\rm{H}}.
\label{BH}
\ee

This black hole solution is an asymptotically AdS black hole which shows a clear van der Waals-Maxwell like phase structure. More discussions about this black hole solution can be found in \cite{phase1}.

We consider a two-side asymptotically AdS black hole coupled with a bath. The existence of the AdS boundary prevents black hole evaporation, so we apply the transparent boundary condition to make the radiation enter the bath. The asymptotically AdS black hole solution is \eqref{s}. As shown in figure 1, we assume that the asymptotically AdS spacetime and the bath are in thermal
equilibrium. We denote $b_{\pm} \left(\pm t_b,r_b\right)$ and $a_{\pm} \left(\pm t_a,r_a\right)$ as the locations of the cutoff surface in the bath and the locations of the island in the asymptotically AdS spacetime, respectively.
\begin{figure}[htbp]
\centering
\subfigure[without island]
{
\begin{minipage}{7cm}
\centering
\includegraphics[scale=0.45]{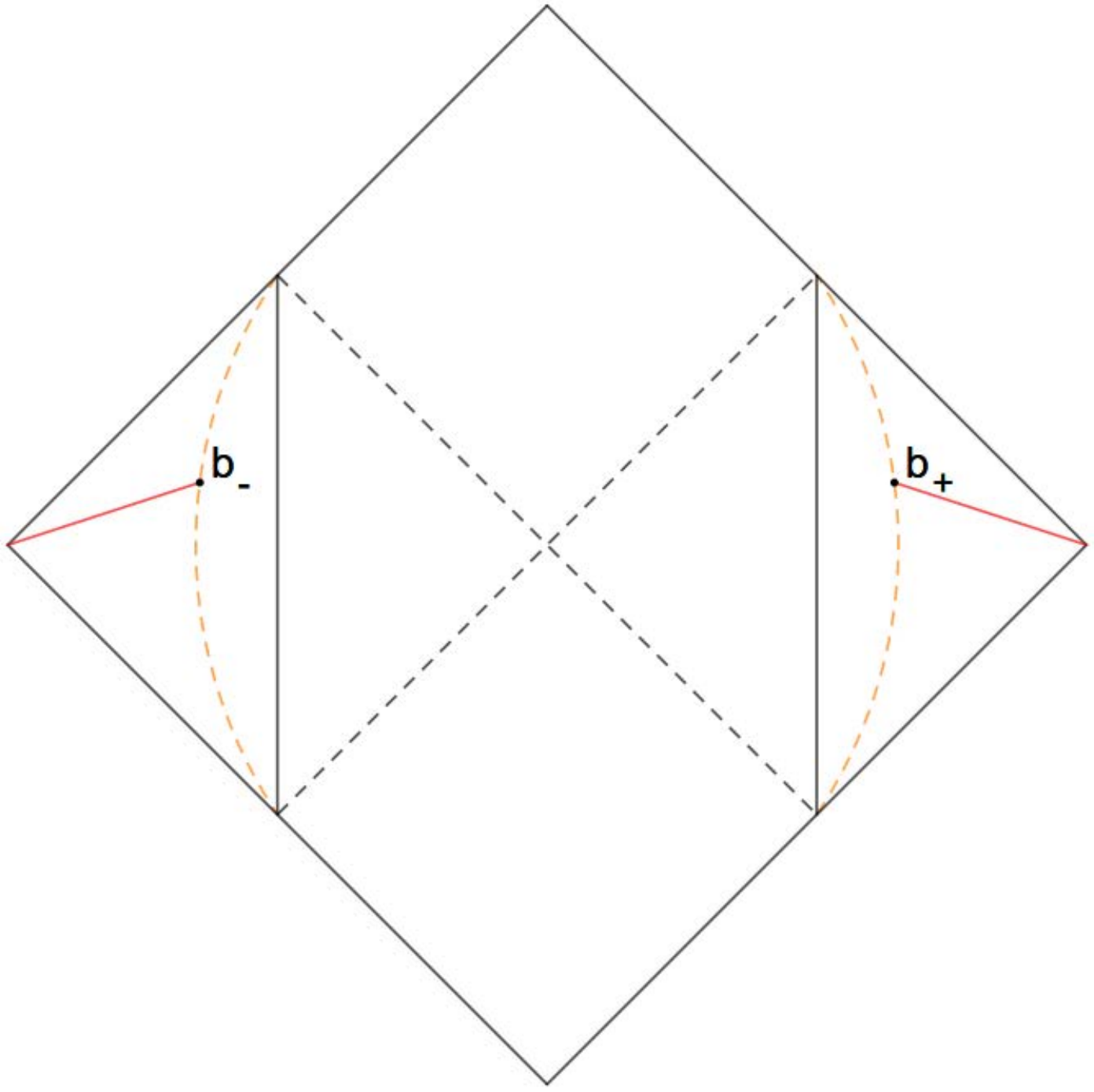}
\label{without island}
\end{minipage}
}
\subfigure[with island]
{
\begin{minipage}{7cm}
\centering
\includegraphics[scale=0.45]{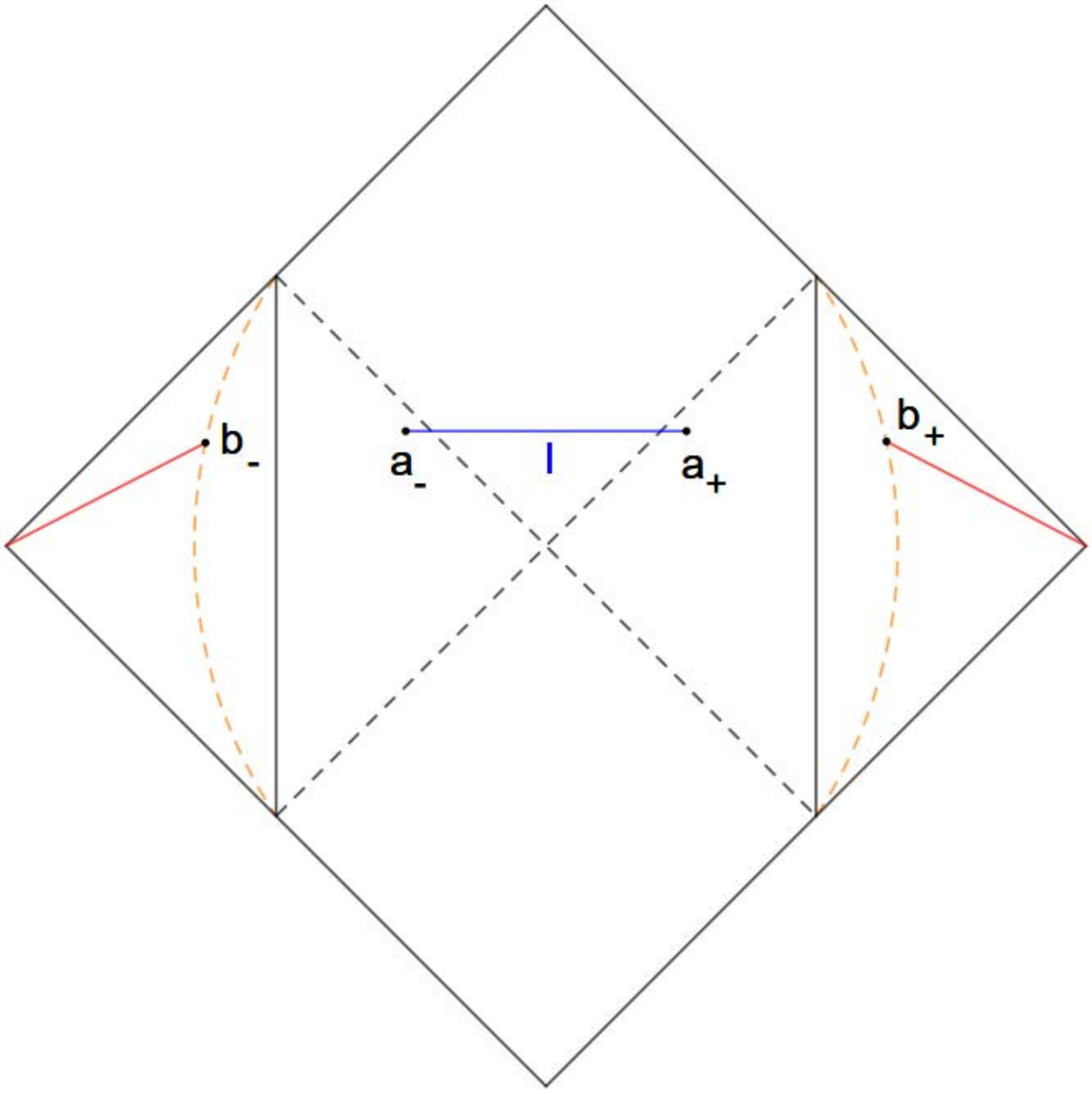}
\label{with island}
\end{minipage}
}
\caption{Penrose diagram for the black hole coupled to a thermal bath. The bath can be viewed as a non-gravitational Minkowski spacetime where $a_{\pm}$ and $b_{\pm}$ correspond to the boundaries of the island and the cutoff surface. }
\label{fig:1}
\end{figure}

In order to obtain the maximal extension of the whole spacetime, we perform the Kruskal transformation, defining the following coordinates,
\be
\begin{aligned}
\text { Right Wedge : } U&=-e^{-\kappa\left(t-r^{*}\right)},&&V=+e^{+\kappa\left(t+r^{*}\right)} , \\
\text { Left Wedge : } U&=+e^{-\kappa\left(t-r^{*}\right)},&&V=-e^{+\kappa\left(t+r^{*}\right)} ,
\label{UV}
\end{aligned}
\ee
and rewrite the metric as
\be
d s^{2}=-\Omega^{-2} d U d V.
\ee
For the metric in \eqref{s}, the twist factor $\Omega$ is given by
\be
\Omega_{\text{2D}}=\frac{\kappa}{\sqrt{ f(r)} } e^{\kappa r^{*}},
\ee
where $\kappa=2\pi T_{\rm{H}}$ is the surface gravity, and $r^{*}=\int f^{-1}(r) d r$ is the tortoise coordinate. For the flat bath region, $f(r)=1$, $r^*=r$, so we can obtain the twist factor in the bath
\be
\Omega_{\text{bath}}=\kappa  e^{\kappa r}.
\label{bath}
\ee

Now, we assume the system is in a vacuum state. So we can use the entanglement entropy formula for 2D vacuum CFT \cite{2.7.}, which reads as
\be
S_{\rm{vN}}=\frac{c}{6} \log L=\frac{c}{6} \log \frac{\left(U_{1}-U_{2}\right)\left(V_{1}-V_{2}\right)}{\Omega_{1}\left(r_1 \right) \Omega_{2}\left(r_2 \right)},
\ee
which represents the entanglement entropy from the contribution of the matter field. Here $L$ is the geodesic distance between two boundary points of a single interval, $\Omega_{1}, \Omega_{2}$ are the twist factors for these two boundary points, $U_{1,2}$ and $V_{1,2}$ are the Kruskal coordinates of these two points in the spacetime. We will use this equation to calculate the entanglement entropy in the following contexts.
\section{Entanglement Entropy }

\qquad In this section, we calculate the entanglement entropy of the radiation region $\left[-\infty, b_{-}\right] \cup\left[b_{+}, \infty\right]$ without the contribution of the island at first. Considering the evolution of the pure state, we only need to calculate the entanglement entropy of the region $\left[b_-,b_+\right]$ because of the complementarity principle, i.e. ,
\be
S_{\rm{rad}}\left(\text{no island}\right) = S_{\rm{vN}}\left[b_-,b_+\right]=\frac{c}{6} \log \frac{\left(U_{b_+}-U_{b_-}\right)\left(V_{b_+}-V_{b_-}\right)}{\Omega_{b_+} \Omega_{b_-}}.
\label{no}
\ee
Inserting $b_+\left(t,r_b\right)$ and $b_-\left(-t,r_b\right)$ into \eqref{UV}, we can obtain the Kruskal coordinates of $b_+$ and $b_-$. As shown in figure \ref{without island}, since coordinates $b_+$ and $b_-$ are in the bath, we can obtain the twist factors by \eqref{bath}, i.e. , $\Omega_{b_+}=\Omega_{b_-}=\kappa  e^{\kappa r_b}$. One inserts the Kruskal coordinates of $b_+$ and $b_-$ into \eqref{no} and obtains
\be
\begin{aligned}
S&=\frac{c}{6} \log \frac{\left(\mathrm{e}^{\kappa(\mathrm{t}+\mathrm{r_b})}+\mathrm{e}^{-\kappa(\mathrm{t}-\mathrm{r_b})}\right)^{2}}{\kappa^{2} e^{2 \kappa r_b}}\\
&=\frac{c}{3} \log \frac{2\mathrm{cosh}\left(\kappa t\right)}{\kappa}.
\end{aligned}
\ee

At late times, the approximation $\kappa t \gg 1$ is valid, so we have
\be
\lim_{t \to \infty}\cosh \left(\kappa t\right) \simeq \frac{1}{2} e^{\kappa t},
\ee
and then obtain,
\be
S \simeq \frac{c}{3} \kappa t.
\label{without}
\ee

Because of the late time approximation, we can find the increase of the entanglement entropy agrees with the original Hawking's result. The entanglement entropy grows approximately linearly with time. Therefore, there is a very sharp paradox here. The entanglement entropy diverging over time violates the unitary principle of quantum mechanics. In order to satisfy the unitary, we introduce the island to avoid such divergence.

\subsection{Cutoff surface far away horizon}
\qquad Now, we consider the cutoff surface that is far away from the horizon first and then consider the cutoff surface that is near the horizon for calculating the entanglement entropy with islands. The island emerges at late times as shown in figure \ref{with island}. At large distances and late times approximation, the generalized entropy can be expressed as twice of the fine-grained entropy of the region $\left[a_+,b_+\right]$ with the contribution of the island \cite{ee},
\be
S_{\rm{gen}}=4\pi r_a+\frac{c}{3} \log \frac{\left(U_{b_+}-U_{a_+}\right)\left(V_{b_+}-V_{a_+}\right)}{\Omega_{b_+} \Omega_{a_+}},
\label{island}
\ee
where the first term is the area term in two-dimensional gravity and the second term represents the contribution of the matter field. $a_+$ and $b_+$ are the locations of the island and the cutoff surface in the right wedge of figure \ref{with island}, respectively. One inserts the coordinates of $b_+\left(t_b,r_b\right)$ and $a_+\left(t_a,r_a\right)$ into \eqref{island}, and obtains
\be
S_{\rm{gen}}=4 \pi r_a+\frac{c}{3} \log \frac{\sqrt{f(r_a)}\left(2 \mathrm{e}^{\kappa\left(r_a^{*}+\mathrm{r_b}\right)} \cosh \left[\kappa\left(t_{b}-t_{a}\right)\right]-\left(\mathrm{e}^{2 \kappa \mathrm{r_a}^{*}}+\mathrm{e}^{2 \kappa \mathrm{r_b}}\right)\right)}{\kappa^{2} e^{\kappa\left(r_a^{*}+r_b\right)}}.
\label{t}
\ee
At first, we extremize $S_{\rm{gen}}$ with respect to $t_a$, i.e.
\be
\frac{\partial S_{\text {gen }}}{\partial t_a}=\frac{c \kappa \sinh[\kappa(t_b-t_a)]}{3(\cosh [\kappa(r_a^*-r_b)]-\cosh [\kappa(t_b-t_a)])}=0.
\ee
The only solution is $t_a=t_b$. Then, \eqref{t} can be rewritten in the following simpler form,
\be
S_{\rm{gen}}=4\pi r_a+\frac{2 c}{3} \log \left[\frac{\mathrm{e}^{\kappa r_b}-\mathrm{e}^{\kappa r_a^{*}}}{\kappa}\right]+\frac{c}{6} \log \left[f\left(r_a\right) \mathrm{e}^{-2 \kappa\left(r_a^{*}+r_b\right)}\right],
\label{S}
\ee
where $r_a, r_b$ are the radial-coordinates of the island and the cutoff surface. To obtain the location of the island, we extremize $S_{\rm{gen}}$ with respect to $r_a$,
\be
\frac{\partial S_{\text {gen }}}{\partial r_a}=4\pi-\frac{2 c}{3} \frac{\kappa}{f\left(r_a\right)\left(\mathrm{e}^{\kappa\left(r_b-r_a^{*}\right)}-1\right)}+\frac{c}{6} \frac {f^{\prime}\left(r_a\right)-2 \kappa}{f\left(r_a\right)}=0.
\label{a}
\ee
Here, we take the near horizon approximation as follows,
\be
r^{*}\approx \frac{1}{2 \kappa} \log \frac{r-r_{\rm{H}}}{r_{\rm{H}}}, \quad
f(r) \approx 2 \kappa\left(r-r_{\rm{H}}\right),
\ee
where the denominator makes the whole formula dimensionless for the tortoise coordinate. By the near horizon approximation, we insert the $f\left(r_a\right)$ approximation into \eqref{a} and  can be simplified as
\be
\frac{12 \pi}{c}=\frac{e^{\kappa r_{a}^{*}}}{\left(r_{a}-r_{\rm{H}}\right)\left(e^{\kappa r_{b}}-e^{\kappa r_{a}^{*}}\right)}-\frac{1}{2} \frac{f^{\prime}\left(r_{a}\right)-f^{\prime}\left(r_{\rm{H}}\right)}{2 \kappa\left(r_{a}-r_{\rm{H}}\right)}.
\label{b}
\ee
Invoking $r_{a}^{*}$ approximation into \eqref{b}, we obtain
\be
\begin{aligned}
\frac{12 \pi}{c}&=\frac{\sqrt{\frac{r_{a}-r_{\rm{H}}}{r_{\rm{H}}}}}{\left(r_{a}-r_{\rm{H}}\right)\left(e^{\kappa r_{b}}-\sqrt{\frac{r_{a}-r_{\rm{H}}}{r_{\rm{H}}}}\right)}-\frac{1}{2} \frac{f^{\prime \prime}\left(r_{\rm{H}}\right)}{2 \kappa}\\
&\simeq \frac{1}{\sqrt{r_{\rm{H}}\left(-r_{\rm{H}}+r_a\right)}\left(e^{\kappa r_b}-\sqrt{\frac{-r_{\rm{H}}+r_a}{r_{\rm{H}}}}\right)},
\end{aligned}
\label{c}
\ee
By solving \eqref{c}, we can obtain the location of islands
\be
r_a=r_{\rm{H}}-\frac{c}{12\pi}+\frac{1}{2} r_{\rm{H}} e^{2 \kappa r_b}-\frac{1}{2} \sqrt{r_{\rm{H}}} e^{ \kappa r_b} \sqrt{-\frac{c}{3\pi}+r_{\rm{H}} e^{2 \kappa r_b}}.
\label{d}
\ee
Simplifying the function $r_a\left(c\right)$ by Taylor's expansion in series of $c$, we obtain
\be
r_a = r_{\rm{H}}+\frac{c^{2}}{144 \pi^{2} r_{\rm{H}}} e^{- 2 \kappa r_b}+{\cal O}\left(c^{3}\right).
\label{e}
\ee

From \eqref{e}, we can see that the location of the boundary of the island is outside the event horizon. We can insert \eqref{e} into \eqref{S} and omit the high order terms. We can obtain the expression of the entanglement entropy,
\be
\begin{aligned}
S&\simeq4 \pi r_{\rm{H}}+\frac{c}{6} \log \frac{2 r_{\rm{H}}}{\kappa^3}+{\cal O}\left(c^{2}\right)\\
&\simeq 2 S_{B H}.
\end{aligned}
\label{2S}
\ee

Because of the emergence of islands, the entanglement entropy does not increase but arrives at a maximal value which is twice the Bekenstein-Hawking entropy. The logarithm term comes from the quantum effect of the matter field. Therefore, for the whole evaporation process, the entanglement entropy increases approximately linearly at the beginning, and then does not increase after reaching the maximum value.
\subsection{Cutoff surface near horizon}
\qquad Now let us discuss the case where the cutoff surface is near the horizon, which is different from the case where the cutoff surface is far from the black hole horizon. We follow the similar logic in \cite{high0}. When it nears the horizon, the generalized entropy is given by
\be
S_{\rm{gen}} \simeq 4 \pi r_a-2 K c \frac{8 \pi r_b}{L^{2}},
\label{near}
\ee
where $K$ is a constant and $L$ is the geodesic distance between the endpoint of the island and the cutoff surface,
\be
\begin{aligned}
L&\simeq\frac{1}{\sqrt{2\kappa}}\int_{r_a}^{r_b} \frac{d r}{\sqrt{r-r_{\mathrm{H}}}}\\
&\simeq\frac{\sqrt{2}}{\sqrt{\kappa}}\left(\sqrt{r_b-r_{\rm{H}}}-\sqrt{r_a-r_{\rm{H}}}\right).
\end{aligned}
\label{L}
\ee

We substitute \eqref{L} into \eqref{near} and extremize $S_{\rm{gen}}$ with respect to $r_a$,
\be
\frac{\partial S_{\text {gen }}}{\partial r_a}=\frac{2 r_{b} c \kappa K}{\left(\sqrt{r_{b}-r_{\rm{H}}}-\sqrt{r_{a}-r_{\rm{H}}}\right)^{3} \sqrt{r_{a}-r_{\rm{H}}}}-1=0.
\ee
where approximations $\sqrt{r_a-r_{\rm{H}}} \ll 1$ and $r_b\simeq r_{\rm{H}} $ are taken into account. The minimization occurs at a smaller solution, satisfying $\sqrt{r_a-r_{\rm{H}}} \ll \sqrt{r_b-r_{\rm{H}}}$, we obtain the location of the island as
\be
r_{a}=r_{\rm{H}}+\frac{\left(2 r_{\rm{H}} c \kappa K\right)^{2}}{\left(r_{b}-r_{\rm{H}}\right)^{3}}.
\label{f}
\ee

Obviously, the location of the boundary of the island is also outside and very close to the event horizon. The result is similar to the situation that the cutoff surface is far from the horizon \eqref{e}.

Inserting \eqref{f} to \eqref{near}, we can obtain the entanglement entropy as
\be
\begin{aligned}
S &\simeq4 \pi r_{\rm{H}}-8 \pi c \kappa K \frac{r_{\rm{H}}}{r_{b}-r_{\rm{H}}}\\&\simeq2S_{\rm{BH}}.
\end{aligned}
\ee

Again, the first term of the entanglement entropy is twice of the Bekenstein-Hawking entropy. The second term is a small number and can be omitted. Because of the emergence of the island, whatever we choose the cutoff surface, the entanglement entropy is about twice the Bekenstein-Hawking entropy.

In summary, the entanglement entropy without islands linear growth diverges in later times leading to the information paradox. However, once the island is considered, the configuration of space-time changes: a region that is not connected with the external radiation region appears inside the black hole, i.e., the entanglement wedge of radiation contains the black hole interior. At last, the entanglement entropy reaches a constant at late times with island. According to the island paradigm, we obtain the unitary result of the evolution of the entanglement entropy.
\section{Page curve and Page time}

\begin{figure}[htbp]
\centering
\includegraphics[scale=0.6]{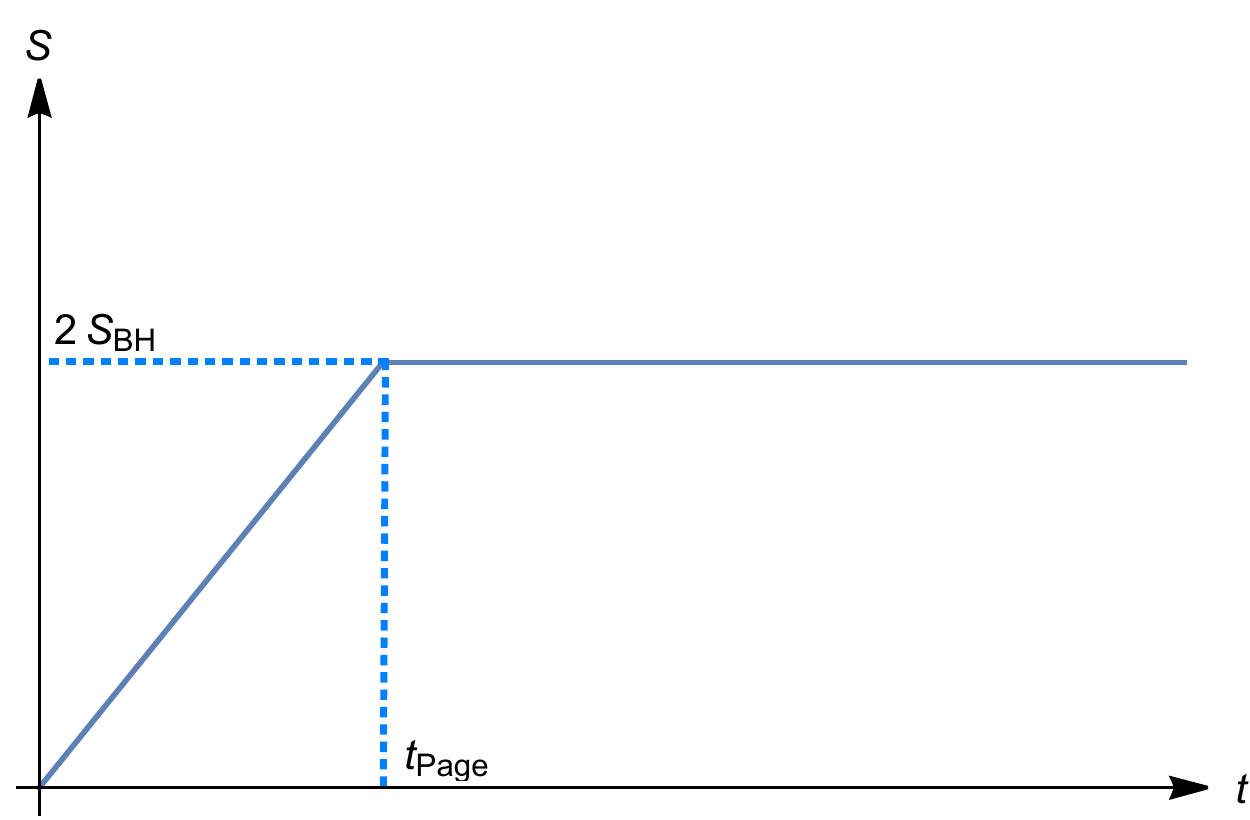}
\caption{Page curve. The entanglement entropy linearly increases at the early time. Then after the Page time, the entanglement entropy arrives at the maximum value -- twice of the Bekenstein-Hawking entropy.}
\label{2}
\end{figure}

\qquad In this section, we obtain the change of the entanglement entropy of the evolution of the black hole. At the early time, from \eqref{without}, the entanglement entropy linearly increases. But because of the island paradigm, after the Page time, the emergence of the island leads to the entanglement entropy does not increase but is a constant \eqref{2S}. Therefore we can obtain the Page curve naturally in figure \ref{2}. By considering \eqref{without} and \eqref{2S}, we obtain the Page time easily,
\be
t_{\text {Page }}=\frac{12 \pi r_{\rm{H}}}{c \kappa}.
\label{page}
\ee

Also, we can obtain the scrambling time which is defined as the minimum time interval in which the information is recovered \cite{s2}. In the island paradigm, this corresponds to the period from the cutoff surface to the boundary of the island. We assume the time interval between sending the information from the cutoff surface $r=r_b$ to the black hole and recovering on the island surface $r=r_a$ to be
\be
\Delta t=r^{*}_{b}-r^{*}_{a}\simeq\frac{1}{2\kappa}\log\frac{r_b-r_{\rm{H}}}{r_a-r_{\rm{H}}}.
\label{scr}
\ee
Substituting the value of $r_a$ from equation \eqref{e} in equation \eqref{scr}, the scrambling time is given by
\be
t_{\rm{scr}} =\frac{1}{2 \kappa} \log \frac{144 \pi^{2}\left(r_{b}-r_{\rm{H}}\right) r_{\rm{H}}}{c^{2} e^{-2 \kappa r_{b}}}\simeq \frac{1}{2 \pi T_{\rm{H}}} \log S_{\rm{BH}},
\label{tscr}
\ee
where we set $r_b$ to have the same order with $r_{\rm{H}}$, $T_{\rm{H}}$ is the Hawking temperature defined \eqref{T}, and $S_{\rm{BH}}=2 \pi r_{\rm{H}}$ is the Bekenstein-Hawking entropy. This result is consistant with \cite{high0,s1}.

\section{Phase transition of the black hole}
\qquad So far, we calculate the change of the entanglement entropy of the whole evaporation process for the special black hole solution and obtain the location of the island and the scrambling time. According to our calculation results, we plot the unitary Page curve. Now we consider the phase transition and analyze the relationship between the Page curve and the phase structure. We consider the charged black hole solution in canonical ensemble with $\eta=1,\zeta=2$, and set the AdS radius $l=1$. Then \eqref{fr} can be simplified as
\be
f(r)=r^2-r^2_{\rm{H}}+a_0 \ln\frac{r}{r_{\rm{H}}}+Q^2\left( r^{-1}-r^{-1}_{\rm{H}}\right).
\ee
For this charged black hole, we can obtain the Hawking temperature and  the entropy from \eqref{T} and \eqref{BH}, which read as
\be
T_{\rm{H}}=\frac{2r_{\rm{H}}^3+a_0 r_{\rm{H}}-Q^2}{4 \pi r_{\rm{H}}^2},\quad S=2 \pi r_{\rm{H}}.
\ee
We also can obtain the expression of the free energy of this black hole solution \cite{phase1},
\be
F=-\frac{1}{2} r_{\rm{H}}^{2}-\frac{Q^{2}}{2} r_{\rm{H}}^{-1}-\frac{a_0}{2}+\frac{a_0}{2} \ln r_{\rm{H}}.
\ee
In the following, we study the chargeless case first and then consider the charged black hole.

\subsection{$Q = 0$}

\qquad For the special case with $Q = 0$, the black hole thermodynamics phase diagram reduces to the Hawking-Page phase transition. We can write down the Hawking temperature and the free energy as
\be
T_{\rm{H}}=\frac{r_{\rm{H}}}{2\pi}+\frac{a_0}{4\pi r_{\rm{H}}},\quad F=-\frac{1}{2} r_{\rm{H}}^{2}-\frac{a_0}{2}+\frac{a_0}{2} \ln r_{\rm{H}}.
\label{TF}
\ee
At the same temperature, for the small black hole, its horizon radius is
\be
r_{s} = \frac{1}{2}\left(2 \pi T_{\rm{H}}-\sqrt{2} \sqrt{-a_0+2 \pi^{2} T_{\rm{H}}^{2}}\right),
\ee
while, for the large black hole, its horizon radius is
\be
r_{l} = \frac{1}{2}\left(2 \pi T_{\rm{H}}+\sqrt{2} \sqrt{-a_0+2 \pi^{2} T_{\rm{H}}^{2}}\right).
\ee
Obviously, there is a minimum temperature $T_{\rm{H}\left(\rm{min}\right)}=\frac{\sqrt{a_0}}{\sqrt{2}\pi}$. For the minimum temperature, two radii are coincident, $r_{l}=r_{s}=\frac{\sqrt{a_0}}{\sqrt{2}}$. Moreover, there are no black hole phases below the minimum temperature.

We can plot the $F$-$T_{\rm{H}}$ diagram as follows.

\begin{figure}[htbp]
\centering
\includegraphics[scale=0.6]{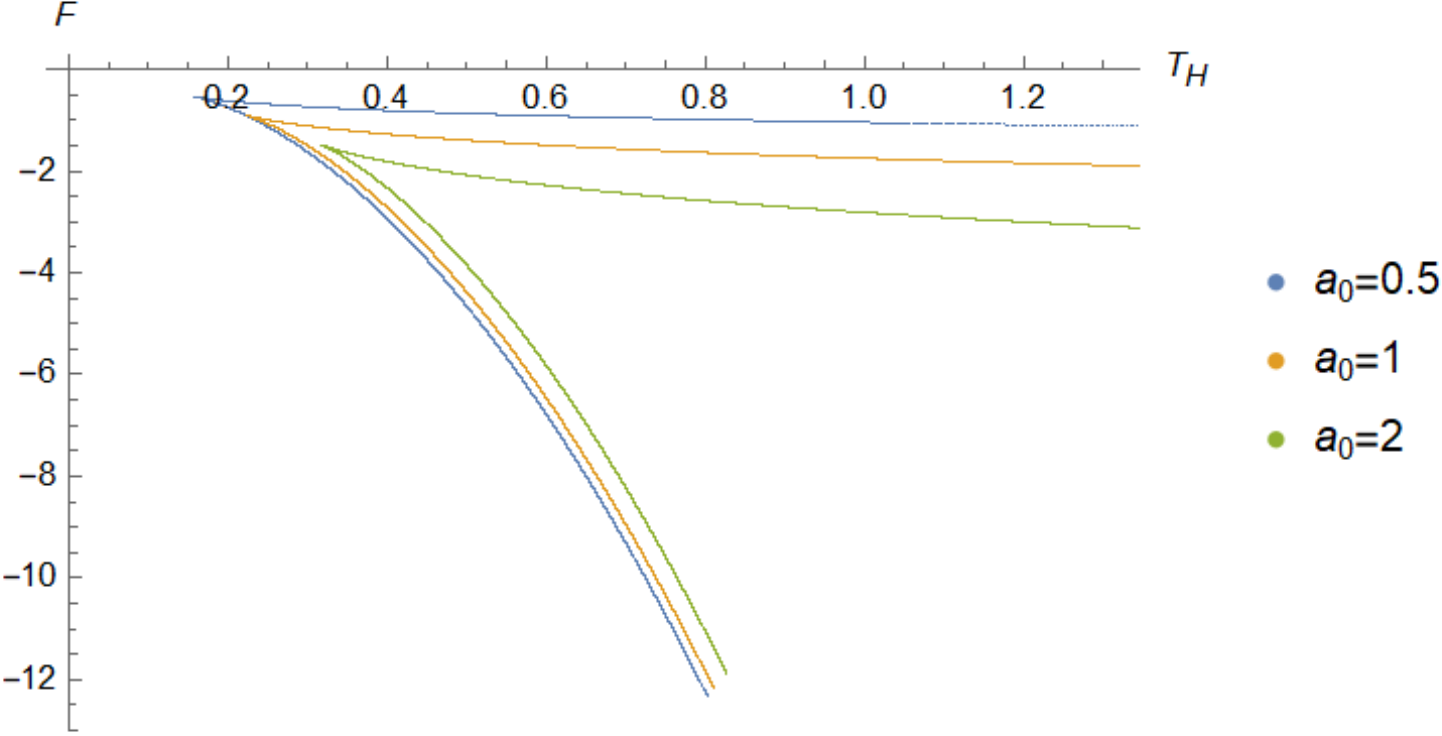}
\caption{The free energy $F$ as a function of temperature $T_{\rm{H}}$ at $a_0=0.5,1,2$. The situation of the Hawking-Page phase structure. The low and high branches correspond to the large and small black holes.}
\label{3}
\end{figure}

As shown in figure \ref{3}, the free energy of the large black hole is always lower than the free energy of the small black hole. Therefore, for the same temperature, the large black hole is the stable phase. Considering the Hawking-Page phase transition, the small black hole is thermodynamically unstable. Therefore, we do not need to consider the island and the Page time of the small black hole, including the extremal black hole.

\begin{figure}[htbp]
\centering
\subfigure[]
{
\begin{minipage}{7cm}
\centering
\includegraphics[scale=0.5]{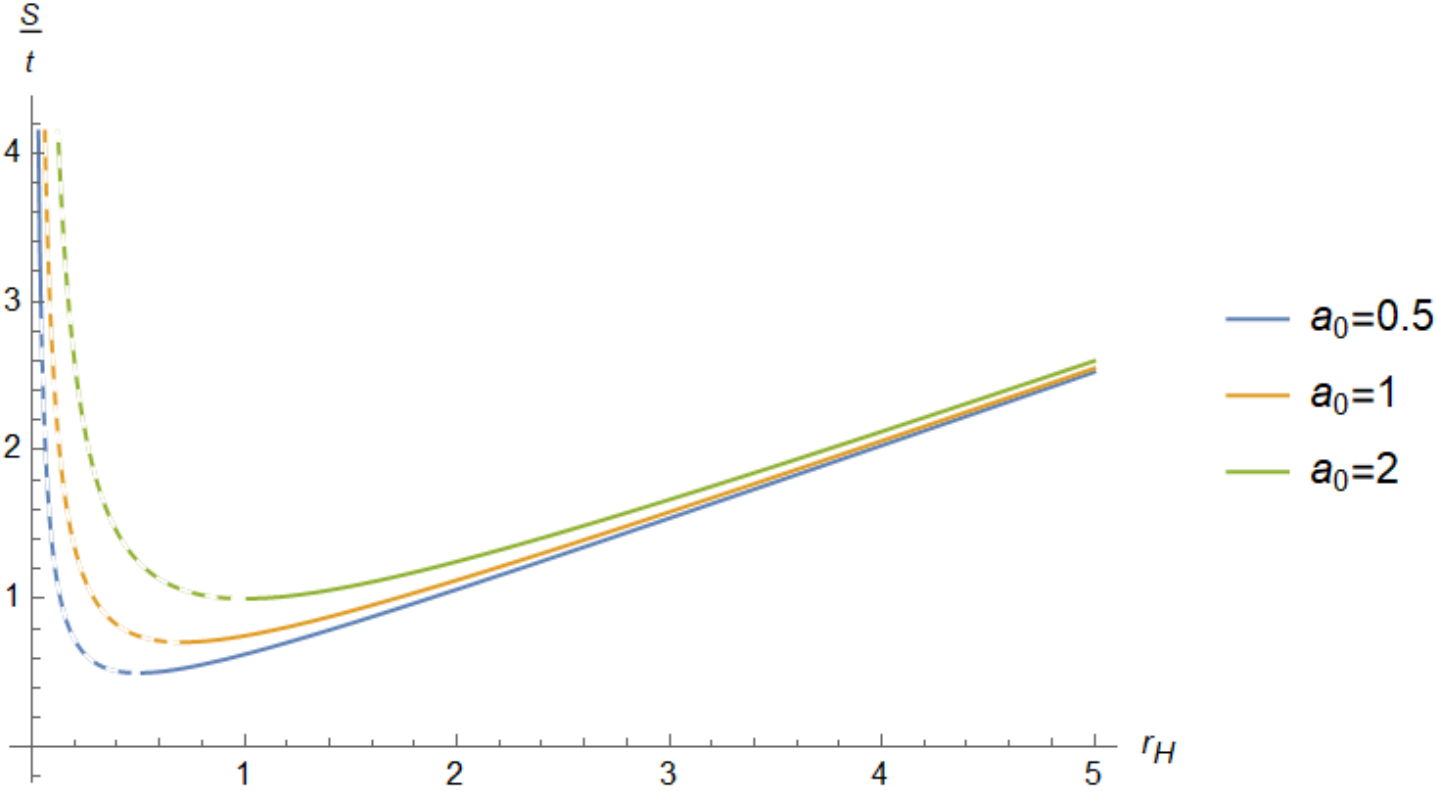}
\end{minipage}
}
\subfigure[]
{
\begin{minipage}{7cm}
\centering
\includegraphics[scale=0.5]{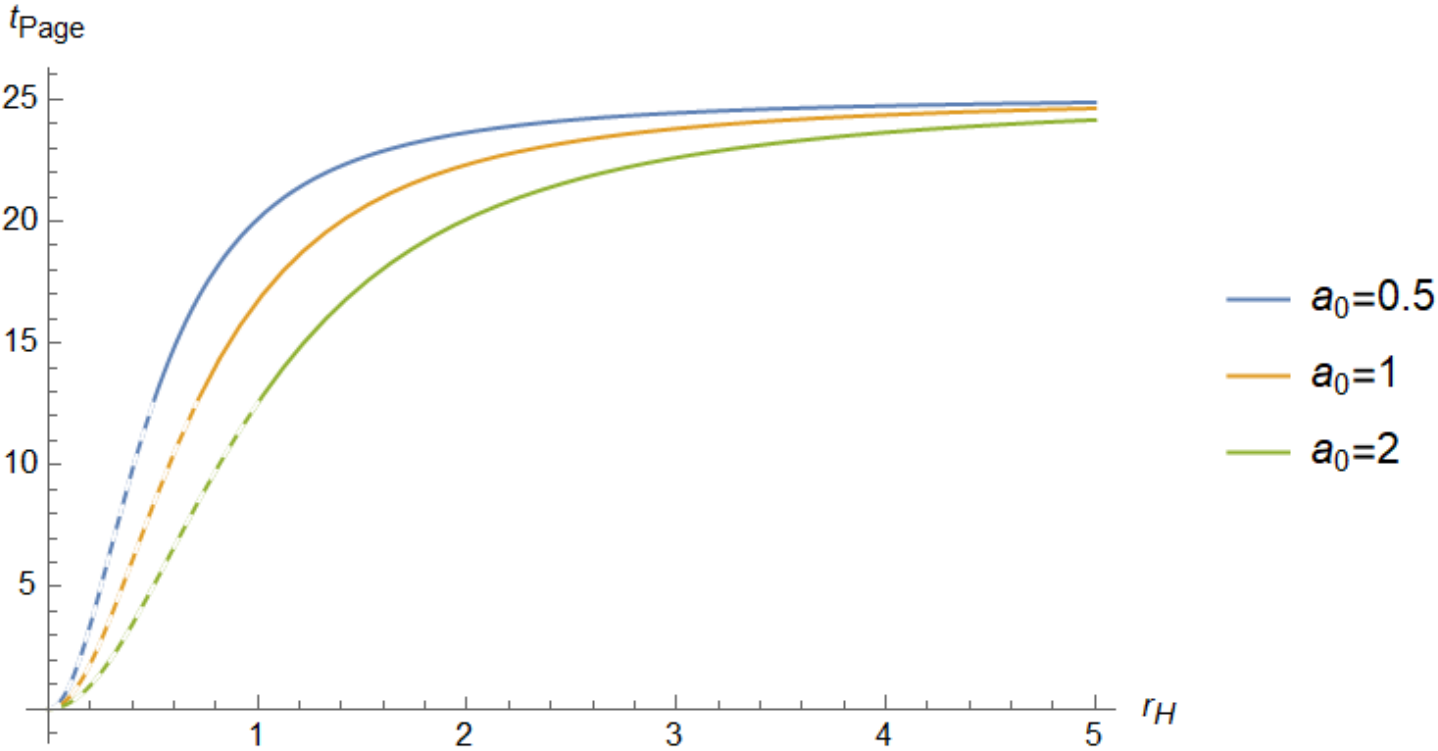}
\end{minipage}
}
\caption{The increased rate of the entanglement entropy and the Page time as a function for the radius at $a_0=0.5,1,2$. The dotted line represents the unstable small black hole.}
\label{4}
\end{figure}

To study the effect on the Page curve from the Hawking-Page phase transition, we consider the situation in the absence of the islands first. From equation \eqref{without}, the entanglement entropy increases linearly at late times. As shown in figure \ref{4}, we obtain the figure for the increased rate of the entanglement entropy and the Page time,
\be
\frac{S}{t}=\frac{c}{3} \kappa.
\ee
We choose $\eta=1,\zeta=2, Q=0, c=1.5$ for \eqref{page} and \eqref{scr}, then we obtain
\be
t_{\text {Page }}=\frac{16 \pi}{2+a_0 r_{\rm{H}}^{-2}}, \quad t_{\text {scr}}=\frac{2 r_{\rm{H}}}{2 r_{\rm{H}}^2+a_0} \log 2 \pi r_{\rm{H}}.
\ee
Now we consider the island paradigm,
\be
S_{\mathrm{rad}} \left(r_{\mathrm{H}},t\right)=\text{Min}\left[S\left(\text{no island}\right),S\left(\text{with island}\right)\right]=\text{Min}\left[\frac{c}{3} \log \frac{2 \cosh (\kappa t)}{\kappa},2S_{\mathrm{BH}}\right].
\label{Srad}
\ee
The whole process of the evolution of the entanglement entropy should satisfy \eqref{Srad}. Considering the Hawking-Page phase transition, we set $c=1.5$ and plot the Page curve as a function of time $t$ and radius $r_{\rm{H}}$ for $a_0=1$ in figure \ref{th}.
\begin{figure}[htbp]
\centering
\includegraphics[scale=0.55]{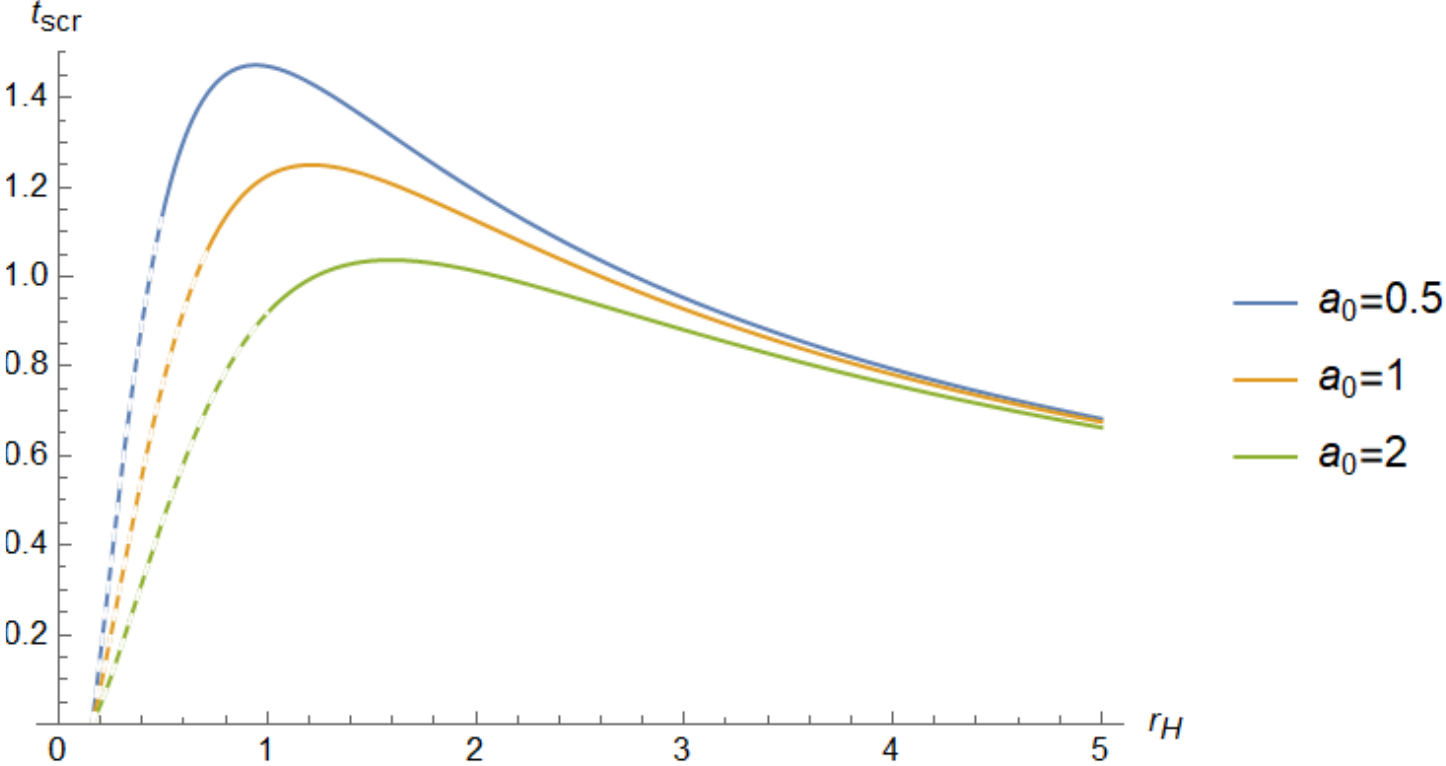}
\caption{The scrambling time as a function for the radius at $a_0=0.5,1,2$. The dotted line represents the unstable small black hole.}
\label{6}
\end{figure}
\begin{figure}[htbp]
\centering
\includegraphics[scale=0.45]{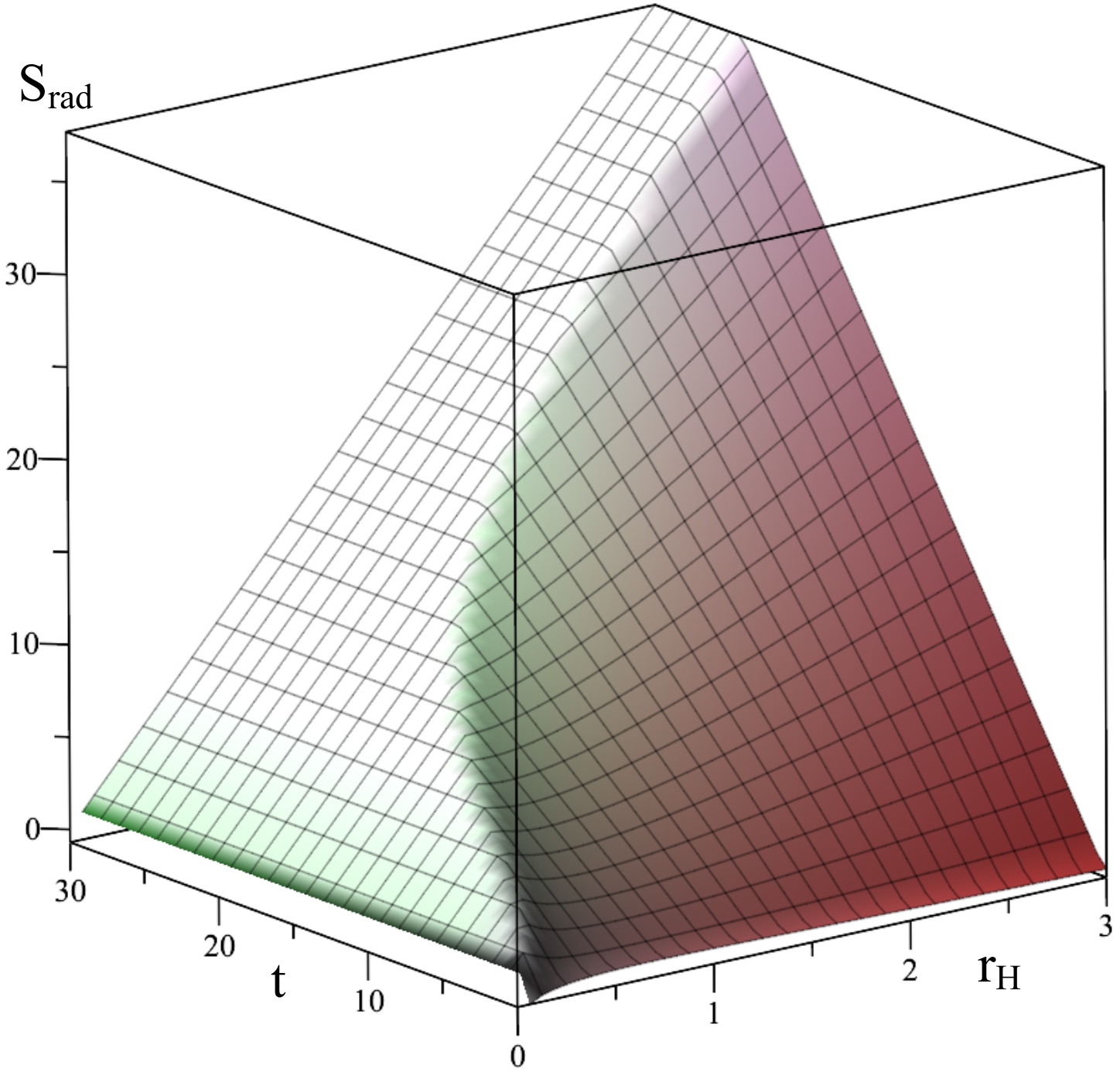}
\caption{Page curves as a function of time $t$ and radius $r_{\rm{H}}$ for $a_0=1$ and $c=1.5$.}
\label{th}
\end{figure}
\subsection{$Q \neq 0$}

\qquad For the general case, we set $\eta=1,\zeta=2,a_0=1,l=1$ and obtain the Hawking temperature and the free energy as
\be
T_{\mathrm{H}}=\frac{2 r_{\mathrm{H}}^3+r_{\mathrm{H}}-Q^2}{4 \pi r_{\mathrm{H}}^2},\quad F=-\frac{1}{2} r_{\mathrm{H}}^2-\frac{Q^2}{2} r_{\mathrm{H}}^{-1}-\frac{1}{2}+\frac{1}{2} \ln r_{\mathrm{H}}.
\label{FT}
\ee
By extremizing $T_{\rm{H}}$ with respect to $r_{\rm{H}}$,
\be
\begin{aligned}
\frac{\partial T_{\rm{H}}}{\partial r_{\rm{H}}}=2 r_{\rm{H}}^3-r_{\rm{H}}+2 Q^2&=0,
\end{aligned}
\label{g}
\ee
we can find only two real, positive solutions
obey \eqref{g}. These two solutions coalesce at the critical point,
\be
\begin{aligned}
\frac{\partial^2 T_{\rm{H}}}{\partial^2 r_{\rm{H}}}=r_{\rm{H}}-3 Q^2=0.
\end{aligned}
\label{h}
\ee
According to equations \eqref{g} and \eqref{h}, we can obtain $r_{\rm{H}\left(\text {crit}\right)}=\frac{1}{\sqrt{6}}$, $Q_{\left(\text {crit}\right)}=\frac{1}{\sqrt{3\sqrt{6}}}$, $T_{\rm{H}\left(\text {crit}\right)}=\frac{1}{2 \pi}\sqrt{\frac{3}{2}}$. Also, we can plot the $F$-$T_{\rm{H}}$ diagram for the charged situation as follows.

\begin{figure}[htbp]
\centering
\subfigure[]
{
\begin{minipage}{7cm}
\centering
\includegraphics[scale=0.5]{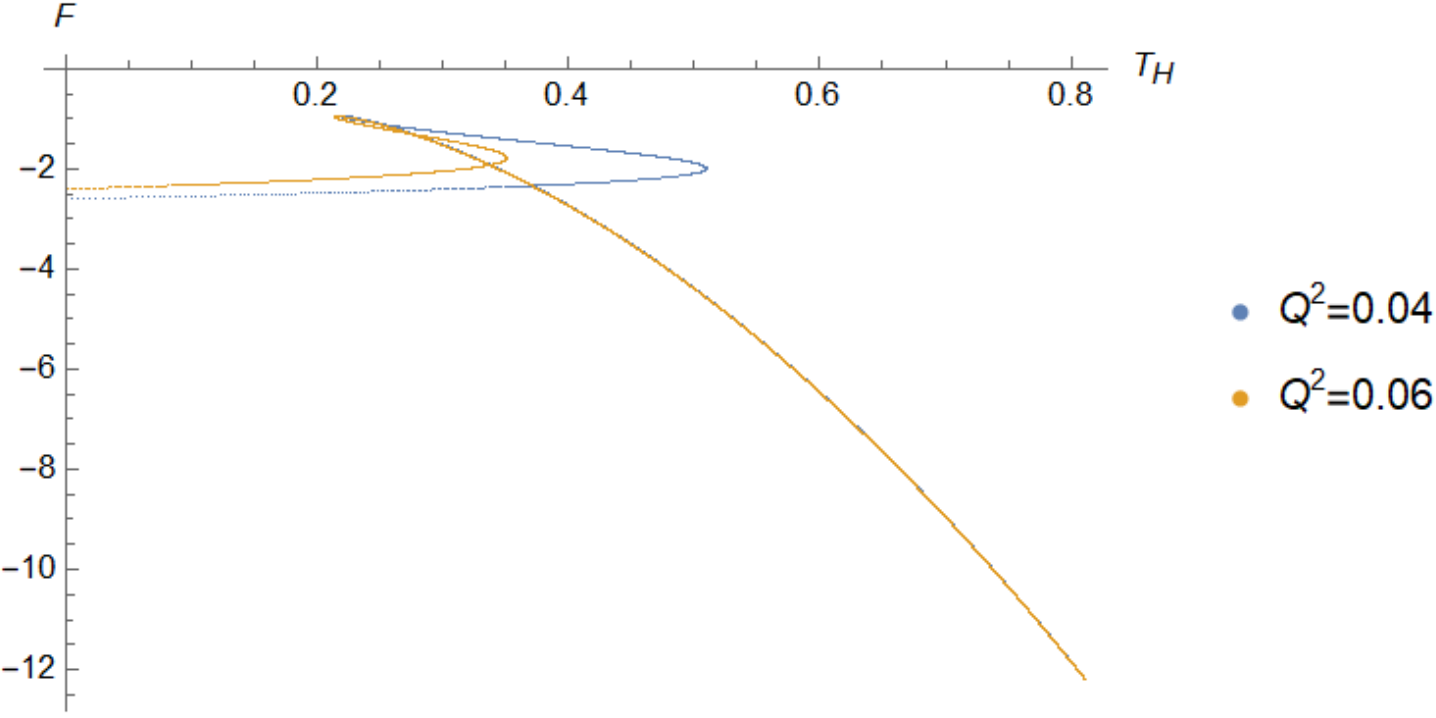}
\label{7a}
\end{minipage}
}
\subfigure[]
{
\begin{minipage}{7cm}
\centering
\includegraphics[scale=0.5]{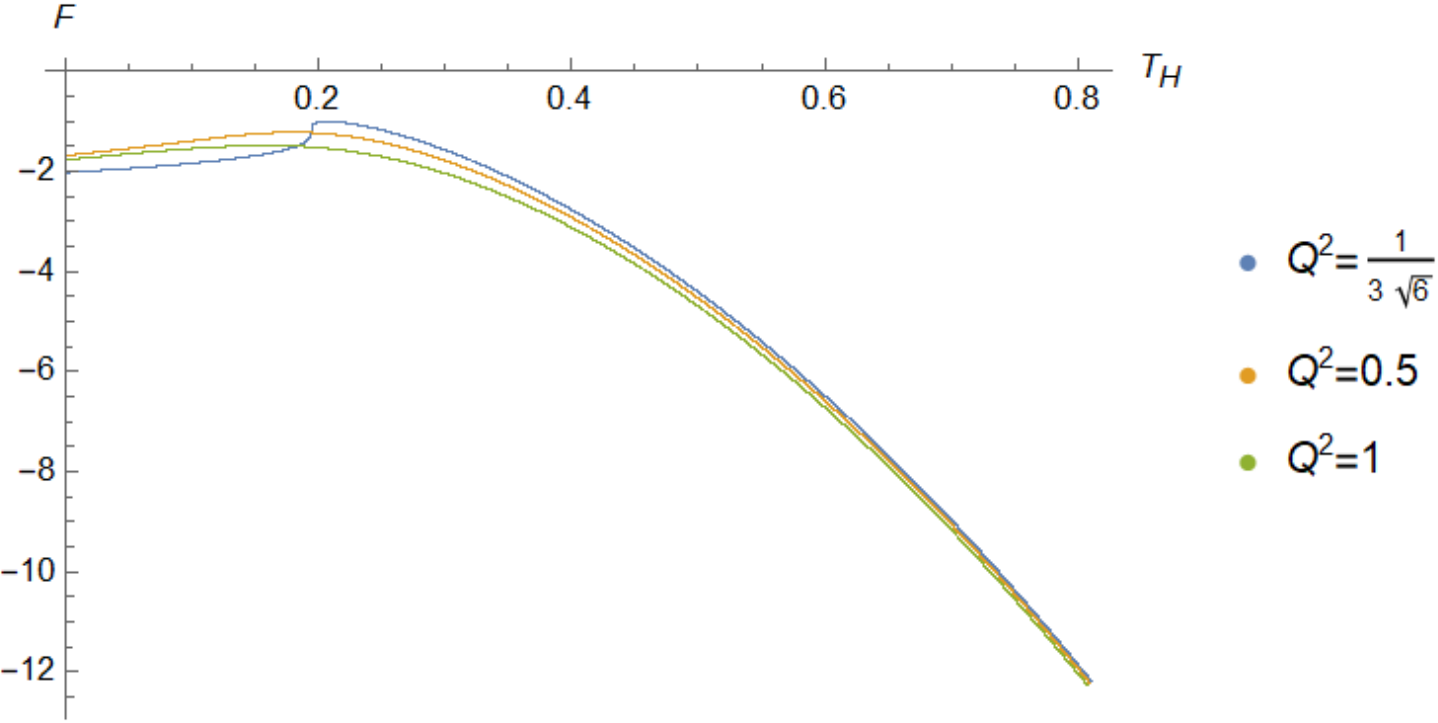}
\label{7b}
\end{minipage}
}
\caption{(a): The free energy $F$ as a function of temperature $T_{\rm{H}}$ at $Q^2=0.04, 0.06$. (b): The free energy $F$ as a function of temperature $T_{\rm{H}}$ at $Q^2=Q^2_{\text { crit }},0.5, 1$.}
\label{7}
\end{figure}

There is a section of the well-known ``swallowtail'' shape in figure \ref{7a}. The intersection points correspond to the phase transition point in the fixed charge ensemble for $Q^2<Q^2_{\rm{crit}}$. The most stable phase requires the lowest free energy. When $Q^2>Q^2_{\rm{crit}}$, the swallowtail structure disappears in figure \ref{7b}, so the phase transition also does not happen.

To investigate the phase transition effects on the Page curve, we plot the increased rate of the entanglement entropy as a function of the radius for several different values of $Q^2$,
\be
\frac{S}{t}=\frac{c}{3} \kappa=\frac{2 r_{\rm{H}}^{3}+r_{\rm{H}}-Q^2}{2 r_{\rm{H}}^{2}}.
\ee

According to \eqref{page} and \eqref{tscr}, we set $c$=1.5 and write down the Page time and the scrambling time versus the radius for any $Q^2$,
\be
\begin{aligned}
t_{\text {Page }}&=\frac{16 \pi}{2+r_{\rm{H}}^{-2}-Q^2 r_{\rm{H}}^{-3}},\\
t_{\text {scr }}&=\frac{2 r_{\rm{H}}^2}{2 r_{\rm{H}}^3+r_{\rm{H}}-Q^2} \log 2 \pi r_{\rm{H}}.
\end{aligned}
\ee

\begin{figure}[htbp]
\centering
\subfigure[]
{
\begin{minipage}{7cm}
\centering
\includegraphics[scale=0.5]{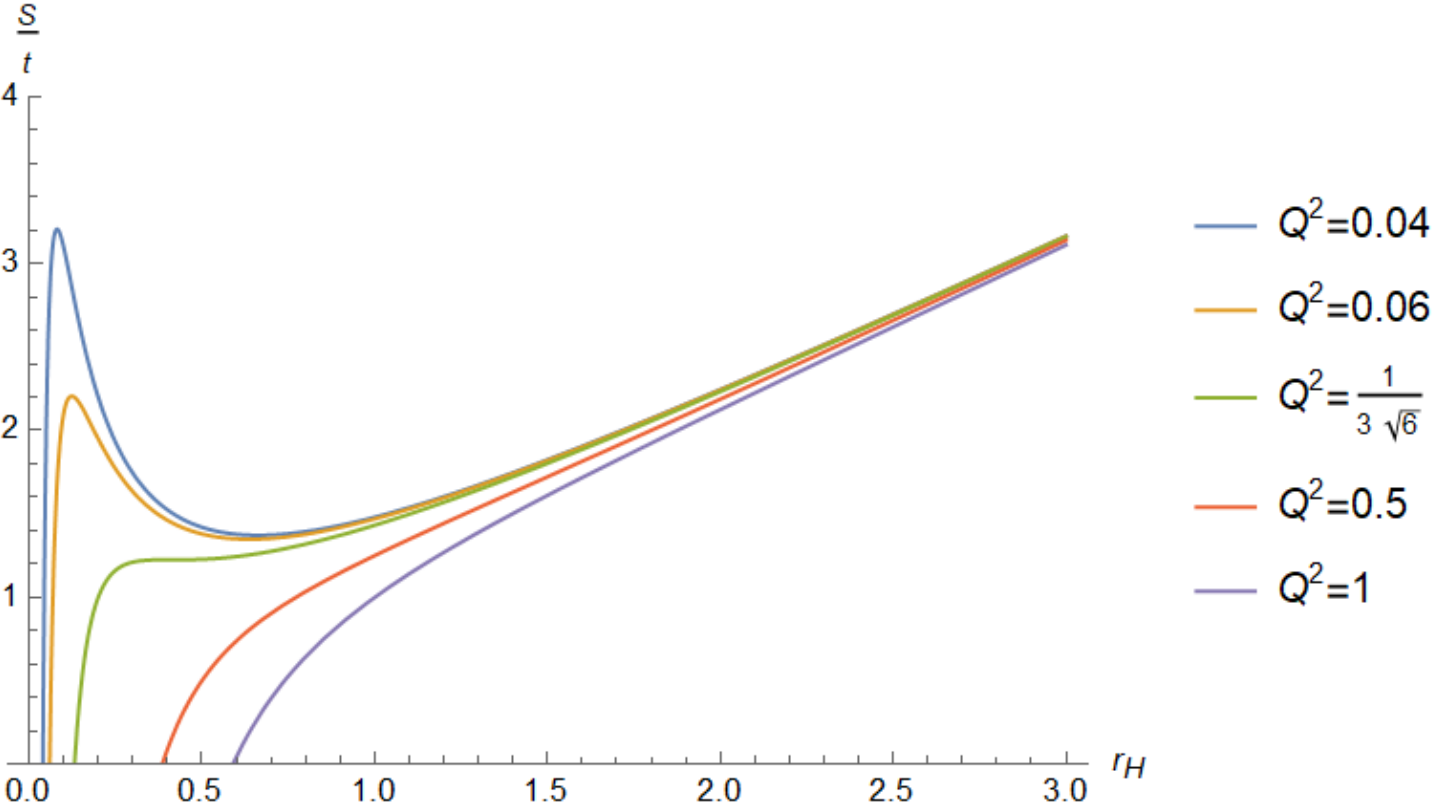}
\label{8a}
\end{minipage}
}
\subfigure[]
{
\begin{minipage}{7cm}
\centering
\includegraphics[scale=0.5]{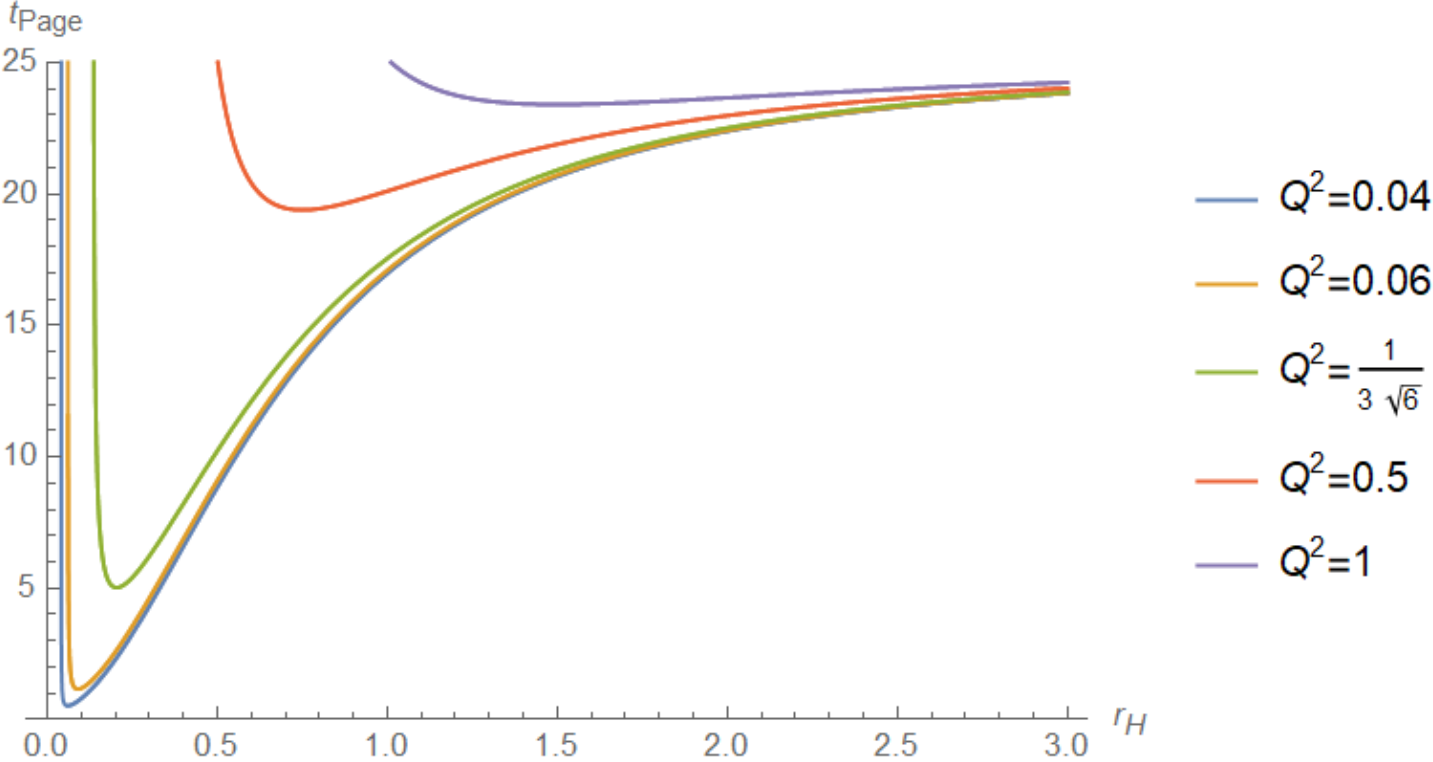}
\label{8b}
\end{minipage}
}
\caption{The increased rate of the entanglement entropy and the Page time as a function for the radius at $Q^2=0.04,0.06,\frac{1}{3\sqrt{6}},0.5,1$.}
\label{8}
\end{figure}

Obviously, in figure \ref{8a}, the increased rate of the entanglement entropy is not a monotonic function of the radius when $Q^2<Q^2_{\rm{crit}}$. The same increased rate corresponds to three different radius values which represent the ``small, middle, large" black hole. However, in real physical processes, the lower the free energy is, the more stable phase is. As figure \ref{7a}, the intersection points are just the critical points where the phase transition happens. Above the critical temperature, the large black hole is more stable, but below the critical temperature, the small black hole is more stable. Thus, figure \ref{7a} behaves like the van der Waals-Maxwell phase structure.

In figure \ref{8b}, it is obvious that the Page time will be a constant $\left(t_{\rm{Page}}=8\pi\right)$, when $r_{\rm{H}}$ approaches infinity. For the large black hole, the charge and entropy do not influence the Page time. However, for the extremal black hole, the Page time is divergent. Since the middle black hole is unstable, so that the physical meaning of the Page time of the middle black hole is not clear when $Q^2<Q^2_{\rm{crit}}$.

\begin{figure}[htbp]
\centering
\includegraphics[scale=0.6]{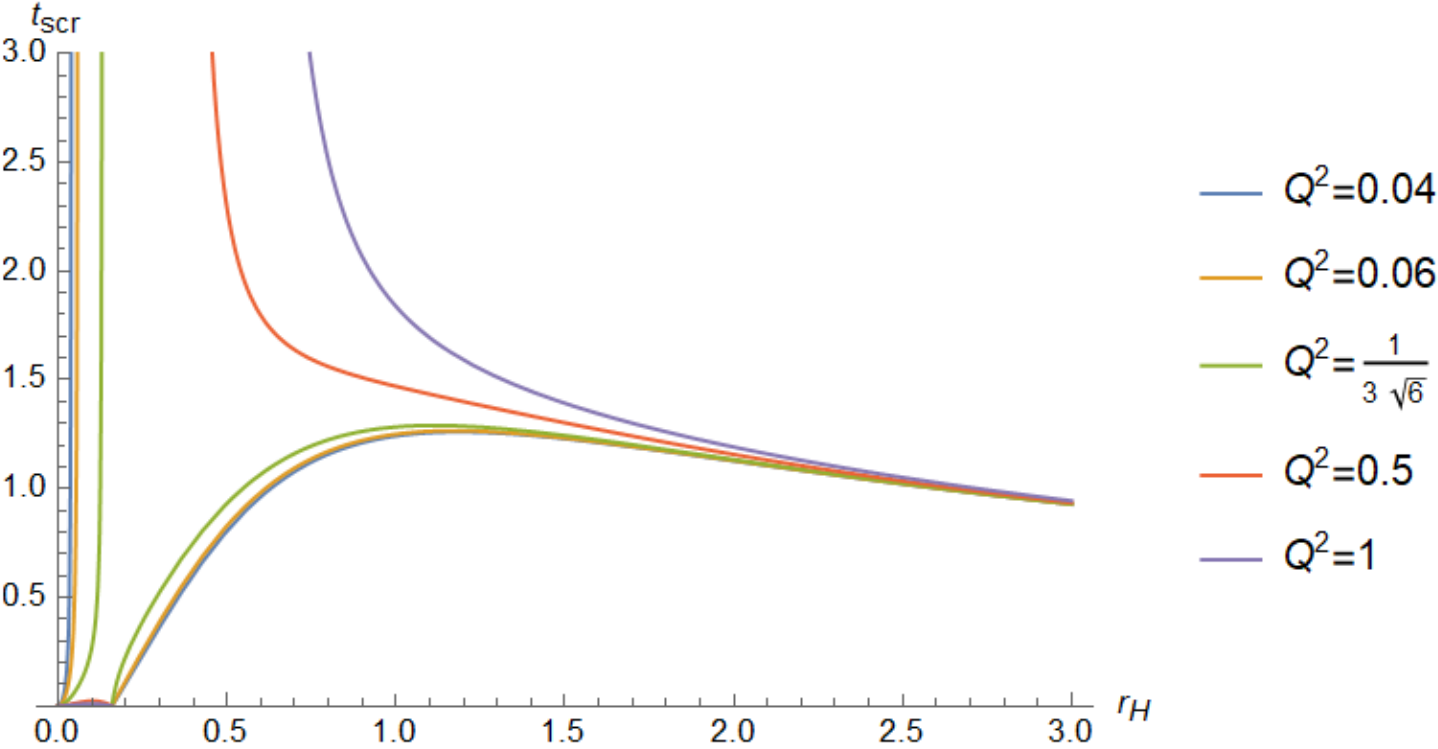}
\caption{The scrambling time as a function for the radius at $Q^2=0.04,0.06,\frac{1}{3\sqrt{6}},0.5,1$.}
\label{fig:2}
\end{figure}

\begin{figure}[htbp]
\centering
\subfigure[]
{
\begin{minipage}{7cm}
\centering
\includegraphics[scale=0.45]{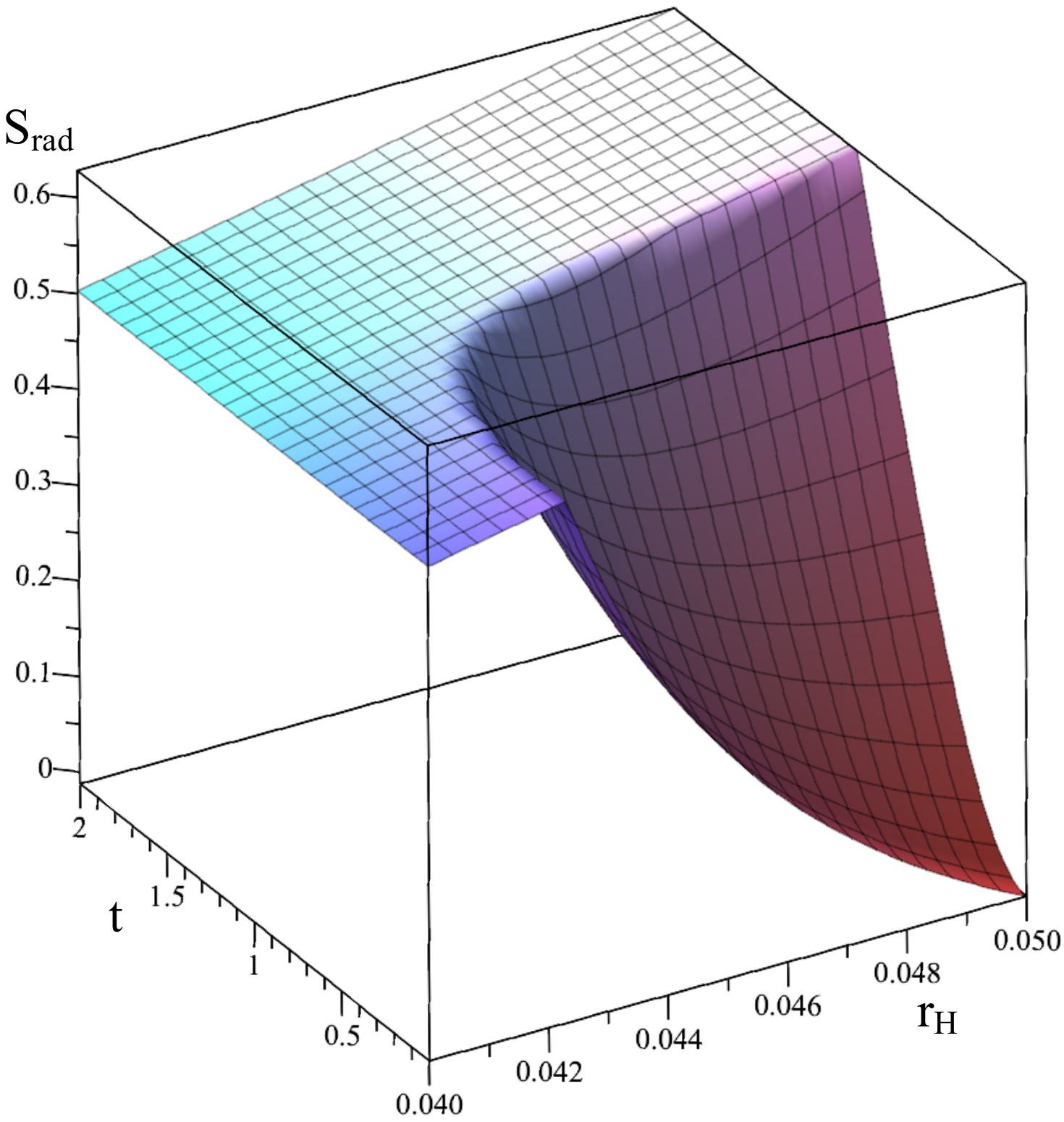}
\label{10a}
\end{minipage}
}
\subfigure[]
{
\begin{minipage}{7cm}
\centering
\includegraphics[scale=0.45]{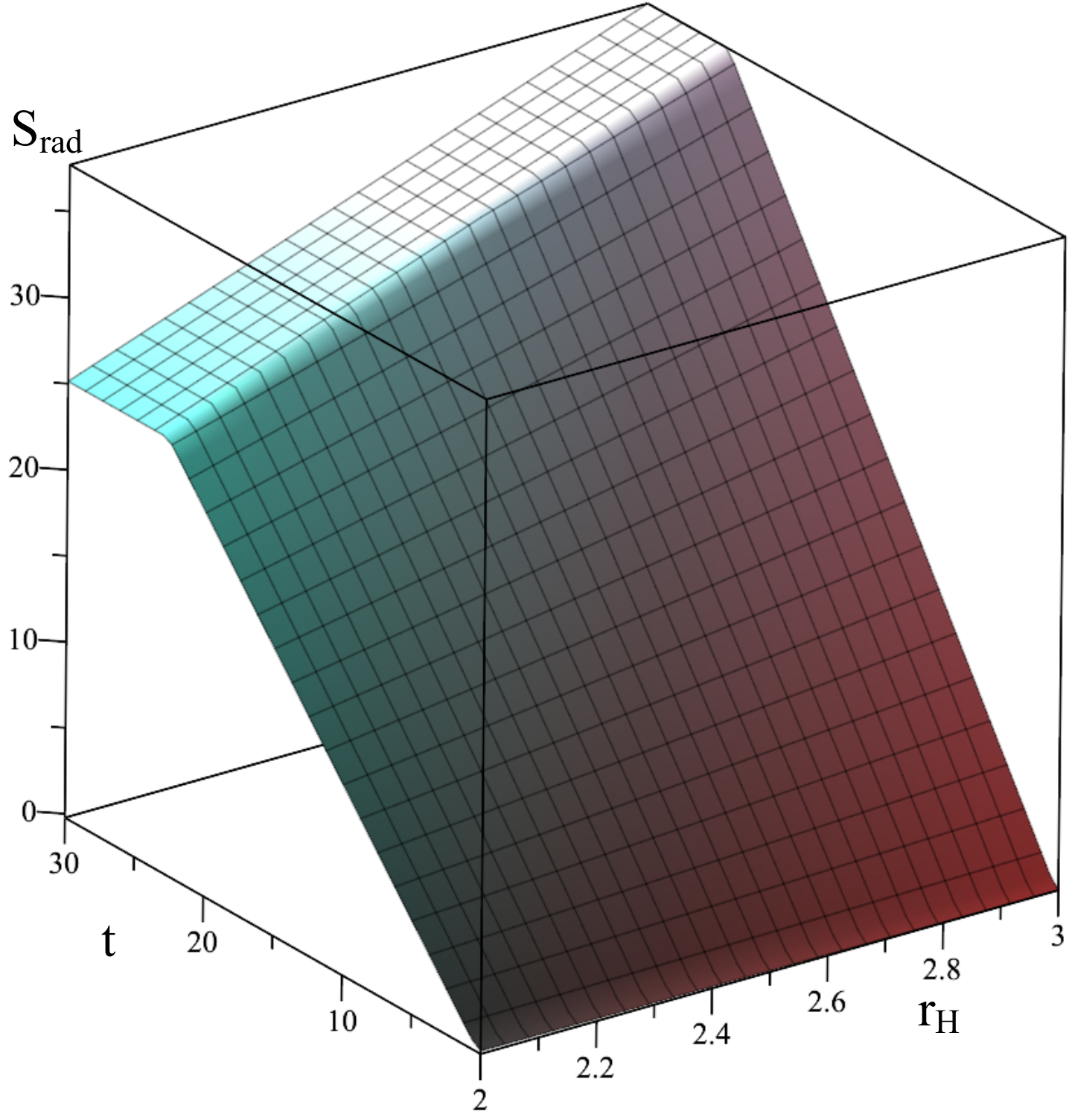}
\label{10b}
\end{minipage}
}
\caption{Page curves for different black holes with $c=1.5$. (a): Page curves as a function of time $t$ and radius $r_{\rm{H}}$ for a series of the small black hole at $Q^2=0.04$. (b): Page curves as a function of time $t$ and radius $r_{\rm{H}}$ for the large black hole at $Q^2=0.04$.}
\label{10}
\end{figure}

Considering the island paradigm \eqref{Srad}, we set $c=1.5$ and plot the Page curve of a series of small and large black holes at $Q^2=0.04$ in figure \ref{10}. As shown in figure \ref{10a}, for the small black hole, the Page time approaches zero. It is different from figure \ref{8b}. Because we take the late time approximation in figure \ref{8b}, it is inaccurate for the Page time of the small black hole. But we do not use the late time approximation in \eqref{Srad}, therefore the result of figure \ref{10} is reasonable.

\begin{figure}[htbp]
\centering
\subfigure[]
{
\begin{minipage}{7cm}
\centering
\includegraphics[scale=0.55]{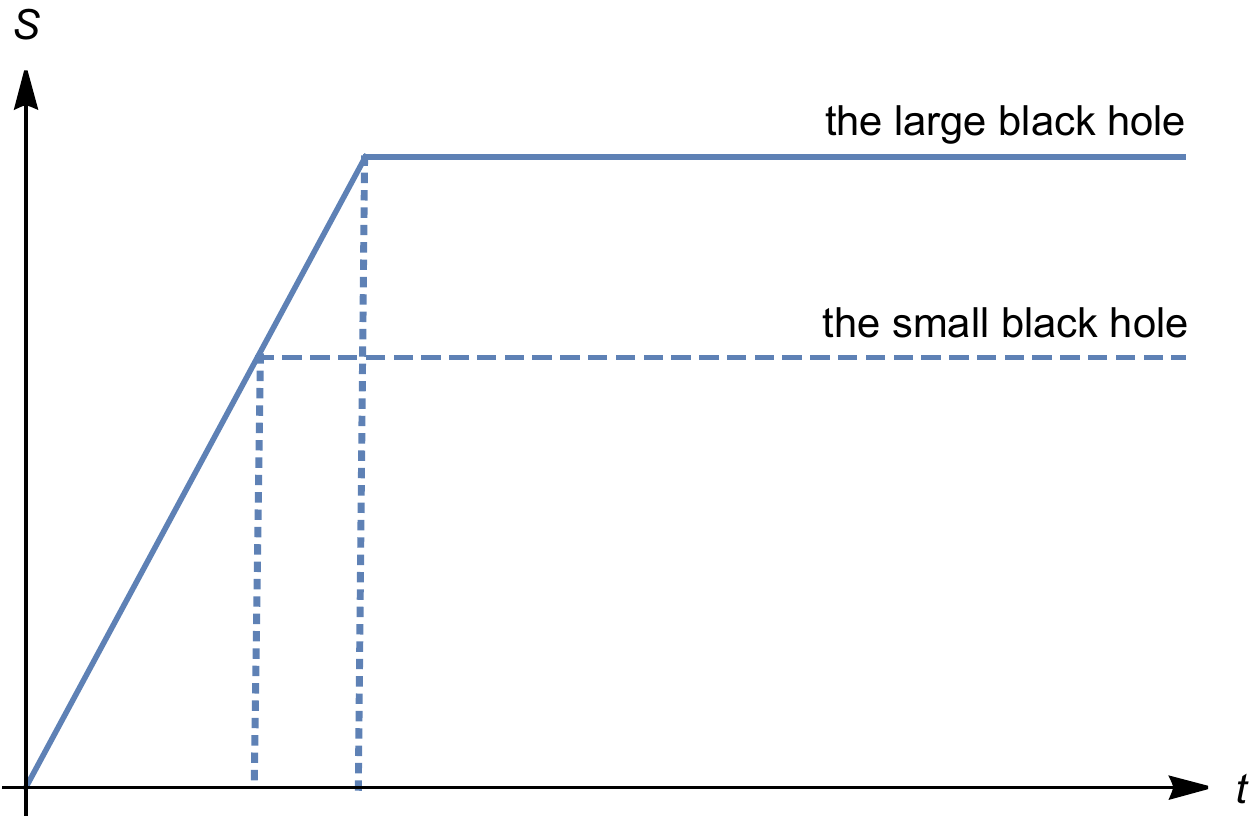}
\label{11a}
\end{minipage}
}
\subfigure[]
{
\begin{minipage}{7cm}
\centering
\includegraphics[scale=0.55]{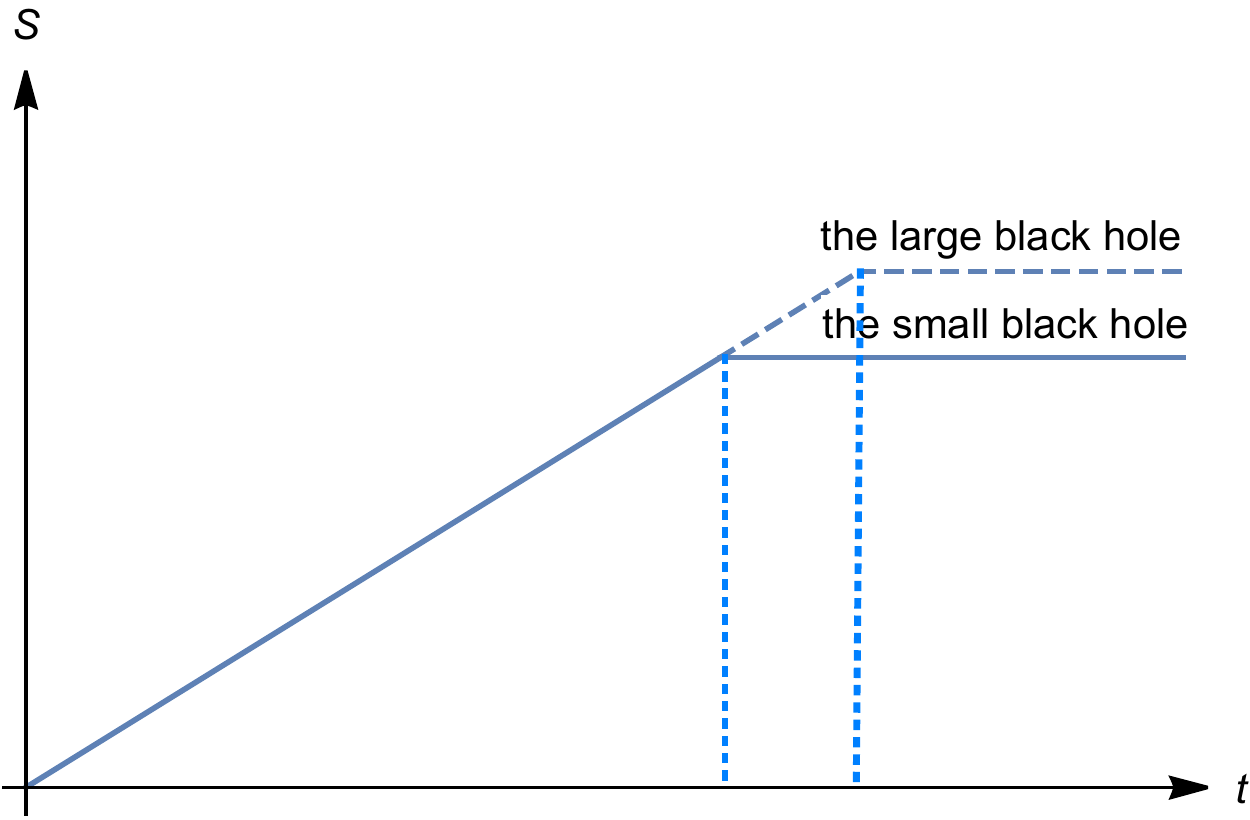}
\label{11b}
\end{minipage}
}
\caption{Page curves for different temperatures. (a): Page curve at the temperature which is higher than the critical temperature. (b): Page curve at the temperature which is lower than the critical temperature.}
\label{11}
\end{figure}

In figure \ref{11}, we plot the Page curve of the black hole with the special phase structure. Above the critical temperature, we can only obtain the Page curve of the large black hole. However, below the critical temperature, we can only obtain the Page curve of the small black hole. The lower the temperature, the greater the Page time. This implies that the lower the temperature, the later the island appears in the black hole's interior.

\begin{figure}[htbp]
\centering
\includegraphics[scale=0.6]{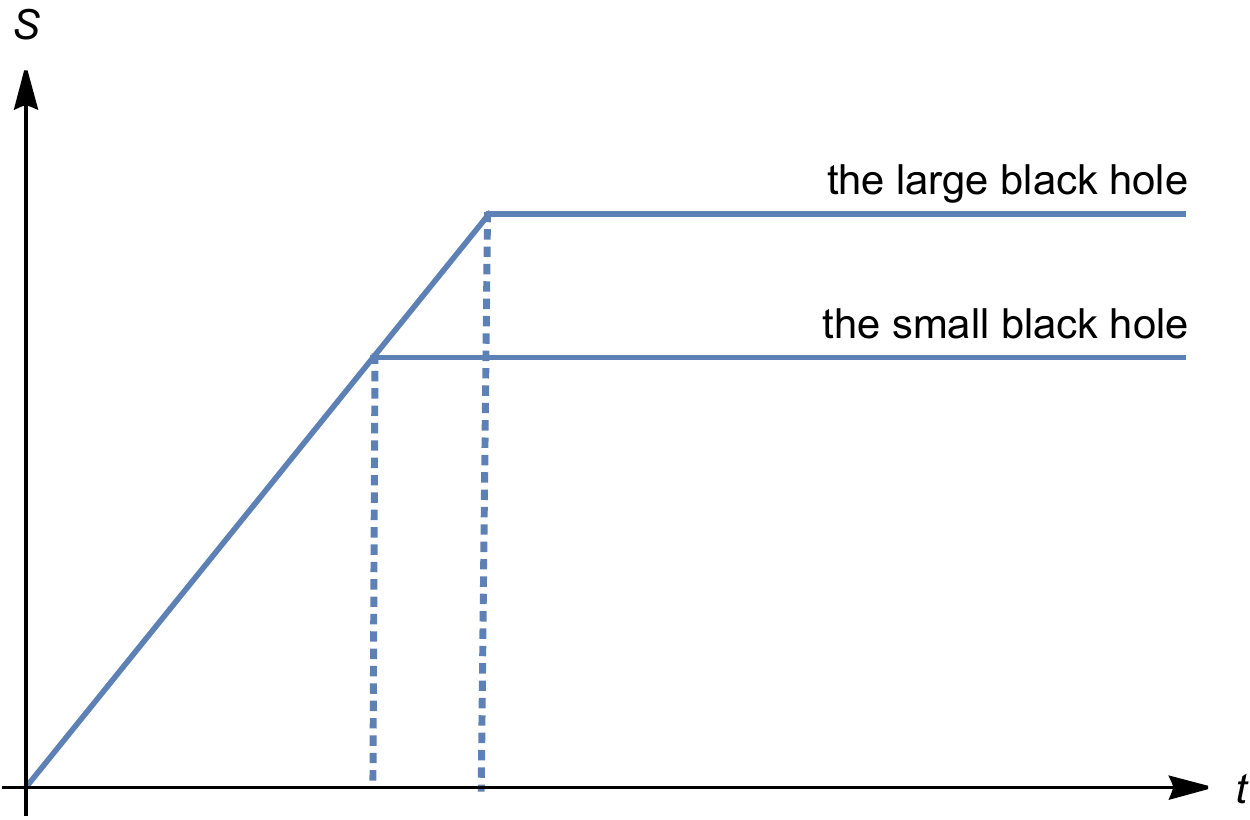}
\caption{Page curve for the critical temperature. The large and small black holes coexist.}
\label{12}
\end{figure}

Taking into account the critical temperature, because of the instability of the middle black hole, the range of the radius of the middle black hole would not become the black hole phase, so the island does not emerge. As the phase transition of the black hole, the location of the island also has a jump. In figure \ref{12}, when the black hole temperature is the critical temperature, the large and small black holes coexist.

\begin{figure}[htbp]
\centering
\includegraphics[scale=0.45]{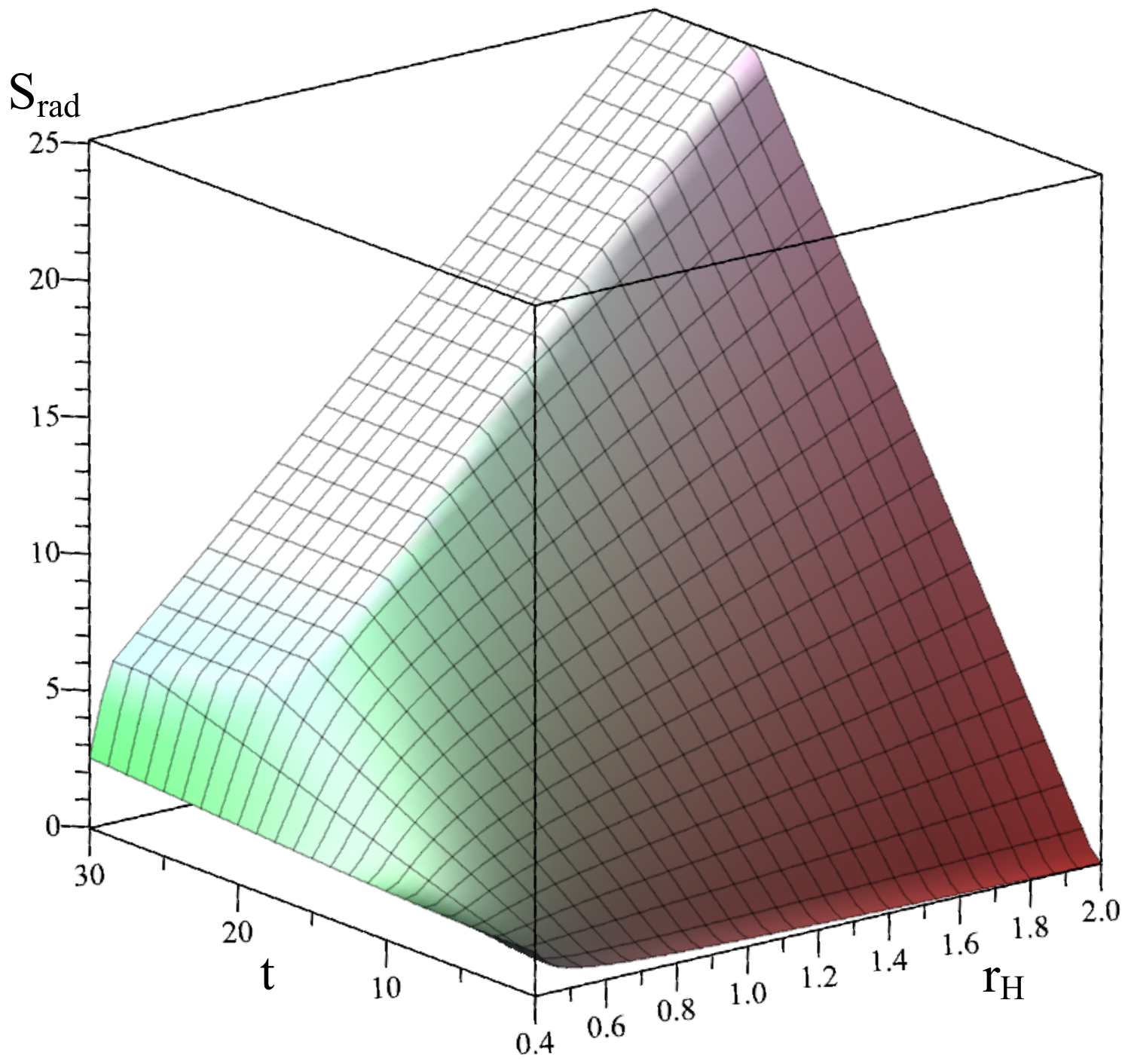}
\caption{Page curves as a function of time $t$ and radius $r_{\rm{H}}$ at $Q^2=0.5$ and $c=1.5$.}
\label{13}
\end{figure}

Finally, we consider the different situations without phase transition when $Q^2>Q^2_{\rm{crit}}$. The green branch corresponds to the critical charge in figure \ref{8}. As shown in figure \ref{8a}, we know the increased rate of the entanglement entropy beyond the critical charge is the monotone function of the radius.  As shown in figure \ref{8b}, with the increase of the charge, the Page time is later. There is no more phase transition, so the Page curve has no special situation for any size black hole just like in figure \ref{2}. We set $c=1.5$ and plot the Page curve as a function of time $t$ and radius $r_{\rm{H}}$ at $Q^2=0.5$ in figure \ref{13}.

 \section{Discussion and Conclusion}

\qquad In summary, we consider the information paradox for a special black hole derived from the deformed JT gravity with the van der Waals-Maxwell like phase structure. To make the black hole evaporate, we construct the eternal black hole by gluing the flat bath. Then we use the island formula to calculate the  entanglement entropy. At early time, the entanglement entropy linearly increases without islands. It matches Hawking's original calculations. At later times, because of the island, the entanglement entropy reaches a constant value which is twice of the Bekenstein-Hawking entropy. Finally, we obtain the scrambling time and the Page curve. We obtain the unitary result and the island outside the event horizon that is consistent with the result of the eternal black hole. Moreover, we consider different situations for the cutoff surface and find the final result is independent of the distance of the cutoff surface.

Moreover, we consider the impact of the phase transition. Phase transition is an important and interesting thermodynamic phenomenon. In fixed charged ensembles with different charges, the black hole has different phase transitions. There is the Hawking-Page phase transition in the black hole without charges. As the charge increases, There is the van der Waals-Maxwell-like phase transition. But if the charge is beyond the critical charge, there are no phase transitions. We study the island and the Page time for the black hole with these phase structures to obtain some results about the influence yielded on the phase transition.

For an neutral black hole without charges, there is only the Hawking-Page phase transition. For the same temperature, there are two black hole phases -- the small and large black holes. By calculating the free energy and the Hawking temperature, we obtain figure \ref{3}, where the large black hole with lower free energy is more stable. Note that the system has a minimum temperature for forming the black hole. So it is meaningless to discuss the island and the Page time below this minimum temperature in this situation. Thus, we only can obtain the Page curve of the large black hole.

For charged black holes, at first, we consider $Q<Q_{\rm{crit}}$, the black hole would have the van der Waals-Maxwell-like phase structure that three different black hole phases may correspond to the same temperature. By calculating the free energy and the Hawking temperature, we obtain figure \ref{7a}, the lowest free energy phase is stable, so only one of the three phases is the stable one. There is a critical temperature where the small black hole phase would jump to the large black hole phase. Of course, the Page time about the middle black hole is meaningless. The situations with the temperature beyond, below, and equal to the critical temperature respectively correspond to figure \ref{11a} which the large black hole is more stable, figure \ref{11b} which the small black hole is more stable, and figure \ref{12} which the small and large black holes coexist. Then we consider $Q\geq Q_{\rm{crit}}$ case in which is the no phase transition happens. The Page time for any temperature is well defined in this situation. Different from the no phase transition situation, the Page time of the small black hole approaches zero in figure \ref{10a}. We do not use the late time approximation to calculate the Page time in \eqref{Srad}. The scrambling time curve monotonically decreases in the no phase transition situation. But in the phase transition situation, there is a maximum scrambling time.

In the doubly-holographic system \cite{KR}, one simply includes a second Karch-Randall brane as a gravitating bath. One finds if the bath itself gravitates, the original Page curve vanishes. And the quantum extremal surface is just at the horizon. These successful works about the introduction of the gravitating bath are in the high dimension \cite{KR,high5,geng}. But there is no graviton in two dimensions. So we can not introduce the KR brane as the gravitating bath in our 2d gravity model. We will consider the reliable construction of the gravitating bath in future work.

In conclusion, we use a simple black hole model with a rich phase structure to compute and analyze the Page curve and study the effect of the phase transition to Page curves. We find the important relationship between the phase transition and the black hole information paradox. We hope that the deep understanding and keen insight of the 2D gravity model in this paper can be applied to the higher-dimensional space-time we live in. As an important phenomenon, the phase transition might be interesting for understanding the information paradox.
\section*{Acknowledgement}
We would like to thank Cheng Ran, Hai-Ming Yuan, Yu-Qi Lei and Qing-Bin Wang for helpful discussions. The study was partially supported by NSFC, China (grant No.11875184 and  No.12275166).

\end{document}